\documentclass[11pt,a4paper]{article}
\pdfoutput=1
\usepackage{a4wide}
\usepackage{amsfonts}
\usepackage{amssymb}
\usepackage{amsmath}
\usepackage{graphicx}
\usepackage{booktabs}
\usepackage{array}
\usepackage{color}
\usepackage{ulem}
\usepackage[small,bf]{caption}
\usepackage{cite}
\usepackage{color}
\usepackage{subfigure}
\def\eps{\epsilon}
\def\1{\mathbf{1}}
\def\3{\mathbf{3}}
\def\2{\mathbf{2}}

\def\e{\varepsilon}

\def\La{\Lambda}
\def\l{\lambda}

\def\th{\theta}

\def\o{\omega}

\def\f{\phi}

\def\ot{\otimes}
\def\op{\oplus}
\newcommand{\zz}{\mathbb{Z}}

\allowdisplaybreaks[1]

\numberwithin{equation}{section}

\newcounter{mysubequation}[equation]

\DeclareMathOperator{\tr}{Tr}

\definecolor{pink}{rgb}{1.,.2,.8}

\newcommand{\bo}[1]{\mathbf{#1}}

\begin{document}

\begin{titlepage}

\vspace*{-15mm}
\begin{flushright}
TTP14-028
\end{flushright}
\vspace*{0.7cm}

\begin{center}
{ \bf\LARGE An SU(5)$\, \times \,$A$_{\mathbf{5}}$ Golden Ratio Flavour Model}
\\[8mm]
Julia Gehrlein \footnote{E-mail: \texttt{julia.gehrlein@student.kit.edu}},
Jens P.\ Oppermann \footnote{E-mail: \texttt{jens.oppermann@student.kit.edu}},
Daniela Sch\"afer \footnote{E-mail: \texttt{daniela.schaefer@student.kit.edu}},
Martin Spinrath \footnote{E-mail: \texttt{martin.spinrath@kit.edu}}
\\[1mm]
\end{center}
\vspace*{0.2cm}
\centerline{\it Institut f\"ur Theoretische Teilchenphysik, Karlsruhe Institute of Technology,}
\centerline{\it Engesserstra\ss{}e 7, D-76131 Karlsruhe, Germany}
\vspace*{1.20cm}

\begin{abstract}
\noindent
In this paper we study an SU(5)$\, \times \,$A$_5$ flavour model which
exhibits a neutrino mass sum rule and golden ratio mixing in the neutrino
sector which is corrected from the charged lepton Yukawa couplings.
We give the full renormalisable superpotential for the model which breaks SU(5) and A$_5$
after integrating out heavy messenger fields and minimising the scalar potential.
The mass sum rule allows for both mass orderings but we will show that
inverted ordering is not valid in this setup. For normal ordering we find the lightest
neutrino to have a mass of about 10-50~meV, and all leptonic mixing angles in
agreement with experiment.
\end{abstract}

\end{titlepage}
\setcounter{footnote}{0}

\section{Introduction}
Experimental results in the lepton sector have shed some new light on the
origin of flavour. In contrast to the quark sector,
lepton mixing angles have the distinctive feature that the atmospheric
angle $\theta_{23}^{\text{PMNS}}$ and the solar angle $\theta_{12}^{\text{PMNS}}$, are both 
rather large \cite{Beringer:1900zz}.  Direct evidence for the reactor angle $\theta_{13}^{\text{PMNS}}$ was
first provided by T2K, MINOS and Double
Chooz~\cite{Abe:2011sj,Adamson:2011qu,Abe:2011fz}. 
Subsequently Daya Bay~\cite{DayaBay}, RENO~\cite{RENO}, and Double
Chooz~\cite{DCt13} Collaborations have measured $\sin^2(2\theta_{13}^{\text{PMNS}})$
to a high precision, see also Tab.~\ref{tab:exp_parameters}.

\begin{table}
\centering
\begin{tabular}{lcc} 
\toprule
Parameter & best-fit ($\pm 1\sigma$) & $ 3\sigma$ range\\ 
\midrule 
$\theta_{12}^{\text{PMNS}}$ in $^{\circ}$ & $ 33.48^{+0.77}_{-0.74}$& $31.30\rightarrow 35.90$\\[0.5 pc]
$\theta_{13}^{\text{PMNS}}$ in $^{\circ}$ & $ 8.52^{+0.20}_{-0.21}$& $7.87\rightarrow 9.11$\\[0.5 pc]
$\theta_{23}^{\text{PMNS}}$ in $^{\circ}$ & $ 42.2^{+0.1}_{-0.1}\oplus 49.4^{+1.6}_{-2.0}$ & $38.4\rightarrow 53.3$\\[0.5 pc]
$\delta_{\text{PMNS}}$ in $^{\circ}$&$251^{+67}_{-59}$&$0\rightarrow 360$\\
\midrule
$\Delta m_{21}^{2}$ in $10^{-5}$~eV$^2$ & $7.50^{+0.19}_{-0.17}$ & $7.03\rightarrow 8.09$\\[0,5 pc]
$\Delta m_{31}^{2}$ in $10^{-3}$~eV$^2$~(NH) &$2.458^{+0.002}_{-0.002}$&$2.325\rightarrow 2.599$\\[0,5 pc]
$\Delta m_{32}^{2}$ in $10^{-3}$~eV$^2$~(IH) &$-2.448^{+0.047}_{-0.047}$&$-2.590\rightarrow -2.307$\\
\bottomrule
\end{tabular}
\caption{The best-fit values and the 3$\sigma$ ranges for the parameters taken from \cite{nu-fit}.
There are two minima for $\theta_{23}^{\text{PMNS}}$. The first one corresponds to the normal hierarchy whereas
the second one corresponds to the inverted hierarchy.
}
\label{tab:exp_parameters}
\end{table}

Among the many proposals trying to address the mixing patterns we will focus here
on models exhibiting the so-called golden ratio (GR) mixing, where $\theta_{12}^{\text{PMNS}}$ is
connected to the golden ratio $\phi_{g}=\tfrac{1+\sqrt{5}}{2}$.

A possible connection was first mentioned as a footnote in \cite{Datta:2003qg} and afterwards implemented
in two different types of golden ratio models. In \cite{Datta:2003qg,Kajiyama:2007gx,Everett:2008et,Cooper:2012bd,Feruglio:2011qq,Albright:2010ap}
they find the prediction $\theta_{12}^{\text{PMNS}}=\tan^{-1}\left(\tfrac{1}{\phi_{g}}\right)\approx 31.7^{\circ}$ (golden ratio type A)
to leading order while in \cite{Adulpravitchai:2009bg,Ding:2011cm,Rodejohann:2008ir,Albright:2010ap} they found
$\theta_{12}^{\text{PMNS}}=\cos^{-1}(\phi_{g}/2)=36^{\circ}$ (golden ratio type B).
 More details on the history
can be found as well in the excellent introduction of \cite{Cooper:2012bd}.
In this work we will find the first relation to leading order. 

The neutrino mixing matrix $U_{\text{GR}}$ will have the form
\begin{equation}
U_{\text{GR}}=
\begin{pmatrix}
	\sqrt{\frac{\phi_{g}}{\sqrt{5}}} & \sqrt{\frac{1}{\phi_{g}\sqrt{5}}} & 0\\
	-\sqrt{\frac{1}{2\phi_{g}\sqrt{5}}} & \sqrt{\frac{\phi_{g}}{2\sqrt{5}}} & \frac{1}{\sqrt{2}}\\
	\sqrt{\frac{1}{2\phi_{g}\sqrt{5}}} & -\sqrt{\frac{\phi_{g}}{2\sqrt{5}}} & \frac{1}{\sqrt{2}}
	\end{pmatrix}
	 P_{0} \;,
	 \label{eq:U_GR}
\end{equation}
which is given in the convention of the Particle Data Group \cite{Beringer:1900zz} with the diagonal
matrix $P_{0}$ = Diag$( \text{exp} (- \tfrac{\text{i} \alpha_{1}}{2}), \text{exp}(- \tfrac{\text{i}\alpha_{2}}{2} ),1 )$
containing the Majorana phases. In \cite{Everett:2008et} and \cite{Kajiyama:2007gx} it was shown that this mixing
pattern can emerge from an A$_5$ family symmetry. Hence, we will adopt here as well an A$_5$ family symmetry.
The mixing pattern which arises in GR type B models can be realised by using a $D_{10}$ symmetry \cite{Adulpravitchai:2009bg}
but will not be discussed here any further.

A$_5$ was utilised as well to construct a four family lepton model \cite{Chen:2010ty} and its double cover
A$^{\prime}_5$ was then used to construct a four family model including quarks \cite{Chen:2011dn} and a flavour model
explaining cosmic-ray anomalies \cite{Hashimoto:2011tn}.

If we assume a diagonal charged lepton basis the physical mixing angles are given as
\begin{align}
 \theta_{12}^{\text{PMNS}} &= \tan^{-1}\left(\frac{1}{\phi_{g}}\right)\approx 31.7^{\circ} \;, 
  \label{eq:A5_Winkel1} \\
 \theta_{13}^{\text{PMNS}} &= 0^{\circ} \;, 
 \label{eq:A5_Winkel2}\\
 \theta_{23}^{\text{PMNS}} &= 45^{\circ} \;.
\label{eq:A5_Winkel3}
\end{align}
Especially, $\theta_{13}^{\text{PMNS}}$ is outside of the 3$\sigma$-range of its experimental value,
cf.\ Tab.~\ref{tab:exp_parameters} and therefore golden ratio mixing can only be a leading order estimate
for the mixing angles which have to be corrected properly.

In \cite{Varzielas:2013hga} an A$_5$ flavour model was proposed which accommodates built in perturbations to golden
ratio mixing which predict correlations between the mixing angles.
In \cite{Cooper:2012bd} corrections to golden ratio mixing were achieved by introducing an additional flavon
which perturbs the structure of the Majorana mass matrix and thereby adjusts the mixing angles to be in agreement
with experimental data.

In our work we will use another approach based on the idea of Grand Unification were such corrections
from the charged lepton sector to the neutrino mixing are well motivated. In such a setup one can expect
$\theta_{12}^e$ to be of the order of the Cabibbo angle $\theta_{C}$ leading to a $\theta_{13}^{\text{PMNS}}$
of a few degrees as we will discuss later in more detail. But due to the precise measurement of the reactor angle
only a few of the vast amount of flavour models are realistic and include Grand Unification
\cite{Antusch:2011qg, Marzocca:2011dh, GUTs, Meroni:2012ty}. Furthermore, we are not aware of any A$_5$
golden ratio GUT model.

To be more precise, the model presented in this paper features SU(5) unification. Hence, we can exploit the
recently proposed new Yukawa coupling relations \cite{Antusch:2009gu, Antusch:2013rxa} which are in very
good agreement with experimental results and are an essential ingredient in an SU(5) GUT context for the prediction
$\theta_{13}^{\text{PMNS}} \approx \theta_C/\sqrt{2} \approx 9^\circ$ \cite{Marzocca:2011dh, Antusch:2011qg, Antusch:2009gu, Antusch:2012fb}.

The corrections from the charged lepton sector are indeed not the only ones which have to be taken into account.
Due to a mass sum rule in the neutrino sector the neutrino spectrum is rather heavy especially
for inverted ordering which will induce large renormalisation group (RG) running effects
that exclude the inverted ordering as we will see. For normal ordering the running is much smaller
but still should be taken into account.

The paper is organised as follows: In section~\ref{sec:model} we will discuss the model including
the symmetry breaking sector and the resulting effective Yukawa and mass matrices.
In section~\ref{sec:Phenomenology} the phenomenological implications of the model are discussed
including RGE effects which rule out the inverted hierarchy neutrino mass pattern.
In section~\ref{sec:Summary} we summarise and conclude and in the appendices we present more
technical details about the family symmetry A$_5$ and the messenger sector of the model.

\section{The model}
\label{sec:model}

In this section we present the SU(5)$\, \times \,$ A$_5$ flavour model before we discuss phenomenological
implications. Our discussion is split into two parts. In the first part we will discuss the sector
responsible for the necessary symmetry breaking of the SU(5) gauge group and the A$_5$ family symmetry.
Then it will become clear why we have arranged for certain flavon alignments when we couple the
symmetry breaking fields to the visible matter sector. Namely, the resulting Yukawa and mass
matrices will give us GR mixing in the neutrino sector and a non-diagonal charged lepton Yukawa
matrix of the desired structure.

\subsection{The symmetry breaking sector}

The symmetry breaking sector can be split into two parts. The first sector contains adjoints
of SU(5) and breaks the GUT gauge symmetry and the second sector contains non-trivial representations of A$_5$
which will break the family symmetry in the desired directions.

\subsubsection{The SU(5) breaking superpotential}

We start our discussion with the more compact SU(5) breaking sector.
The GUT group is broken by the vacuum expectation values (vevs) of the two adjoint fields
$H_{24}$ and $H'_{24}$. The field $H_{24}$ will couple to the matter sector
resulting in non-trivial Clebsch-Gordan (CG) coefficients and hence non-standard GUT scale Yukawa coupling
ratios. The superpotential for the adjoint fields reads
\begin{equation}
 \mathcal{W}_{24} = M_{24} \tr H_{24} H'_{24} + \lambda_H \tr (H'_{24})^3 + \lambda_S S^3 + \kappa \, S \tr H_{24}^2 \;,
\end{equation}
where we have also introduced a singlet field $S$. The scalar potential is minimised
by the vevs
\begin{align}
\langle H'_{24} \rangle &= V'_{24} \text{ Diag} \left( 1,1,1, -\tfrac{3}{2}, -\tfrac{3}{2} \right) \;,\\
\langle H_{24} \rangle &= V_{24} \text{ Diag} \left( 1,1,1, -\tfrac{3}{2}, -\tfrac{3}{2} \right) \;,\\
\langle S \rangle &= V_{S} \;,
\end{align}
which fulfil the relations
\begin{align}
( V_{24} )^3 = \frac{1}{15} \frac{\lambda_S}{\kappa^3 \lambda_H} M^3 \;, \quad
( V'_{24} )^2 =\frac{2}{3} \frac{M V_{24}}{\lambda_H} \;, \quad
( V_S )^2 =  \frac{5}{2} \frac{\kappa}{\lambda_S} ( V_{24} )^2 \;.
\end{align}
The vevs of the adjoints break SU(5) to the Standard Model gauge group SU(3)$_C \times$SU(2)$_L \times$ U(1)$_Y$.

The above mentioned superpotential is a modified, combined version of superpotential
(b) and (c) of \cite{Antusch:2014poa} extended by a singlet. In that work the so-called double missing partner
mechanism --- a possible solution to the doublet-triplet-splitting problem in these kind of models --- was discussed.
This mechanism could be applied here as well but the construction of the full
potential goes beyond the scope of the current work.

\subsubsection{The flavon alignment}
\label{sec:alignment}

Now we turn to the flavon alignment sector. Before we discuss the corresponding superpotentials
we want to first give an overview of all the flavons and their alignments. First of all,
there are a couple of flavons which transform as one-dimensional representations under A$_5$
\begin{equation}
 \label{eq:vev1}
 \langle \theta_i \rangle = v_{\theta_i} \;, \;\; i = \text{1, 2, 3} \;, \quad \langle \epsilon_j \rangle = v_{\epsilon_j} \;, \;\; j = 1, \ldots, 5 \;.
\end{equation}
Then we have two flavons in three-dimensional representations
\begin{equation}
 \label{eq:vev3}
 \langle \f_2 \rangle = v_\f^{(2)} \left(0,1,0\right) \;, \quad
 \langle \f_3 \rangle = v_\f^{(3)} \left(0,0,1\right) \;,
\end{equation}
two flavons in five-dimensional representations
\begin{equation}
 \label{eq:vev5}
 \langle \omega  \rangle = \left(\sqrt{\tfrac{2}{3}} (v_2 + v_3), v_3, v_2, v_2, v_3\right) \;,
 \quad  \langle \tilde{\omega}   \rangle = v_1 \left(1,0,0,0,0\right) \;,
\end{equation}
and one flavon in a four-dimensional representation of A$_5$
\begin{equation}
 \label{eq:vev4}
 \langle \l   \rangle = v_\l \left(1,1,1,1\right) \;.
\end{equation}

The alignment for the four- and five-dimensional flavon fields closely
resembles the alignment in \cite{Cooper:2012bd} and \cite{Ding:2011cm}
and hence we will not discuss it here in detail. The superpotential for them
reads
\begin{equation}
\label{eq:sp5}
\mathcal{W}_f = g_1 \o \l D_{\o} + g_2 \l^2 D_\l + g_3 \tilde{\omega}^2 D_{\tilde{\omega}} \;.
\end{equation}
For the three-dimensional flavons the superpotential is of the form
\begin{equation}
\label{eq:sp3}
\mathcal{W}_t = g_4 \f_2\tilde{\o} D_{\o\f}^{(2)} + g_5 \f_3\tilde{\o} D_{\o\f}^{(3)} + g_6
\f_2^2 D_\f^{(2)} + g_7\f_3^2 D_\f^{(3)} \;,
\end{equation}
which upon inserting $\langle \tilde{\omega} \rangle$ yields the non-trivial F-terms
\begin{align}
\label{eq:ft3}
\frac{\partial \mathcal{W}_t}{\partial D_{\o\f,1}^{(2)}} &= \sqrt{3}g_4v_1
\f_{2,1} \;,\\
\frac{\partial \mathcal{W}_t}{\partial D_{\o\f,1}^{(3)}} &= \sqrt{3}g_5v_1
\f_{3,1} \;,\\
\frac{\partial \mathcal{W}_t}{\partial D_{\f}^{(2)}} &= 2g_6
\f_{2,2}\f_{2,3} \;,\\
\frac{\partial \mathcal{W}_t}{\partial D_{\f}^{(3)}} &= 2g_7
\f_{3,2}\f_{3,3} \;.
\end{align}
It is easy to see that these terms vanish given the alignments in eq.~\eqref{eq:vev3}.

Finally, for the one-dimensional flavons we have used the mechanism described
in \cite{Antusch:2011sx, Antusch:2014poa}. The superpotential reads
\begin{align}
\mathcal{W}_s =~&
P\left(\frac{\th_1^6}{\La^{4}} - M^{2}\right) +
P\left(\frac{\th_2^{12}}{\La^{10}} - M^{2}\right) +
P\left(\frac{\th_3^{12}}{\La^{10}} - M^{2}\right) + \nonumber \\
& P\left(\frac{\e_1^3}{\La^{1}} - M^{2}\right) +
P\left(\frac{\e_2^{12}}{\La^{10}} - M^{2}\right) +
P\left(\frac{\e_3^6}{\La^{4}} - M^{2}\right) + \nonumber \\
& P\left(\frac{\e_4^{12}}{\La^{10}} - M^{2}\right) +
P\left(\frac{\e_5^{12}}{\La^{10}} - M^{2}\right) + 
\mathcal{O}(P^3) \;,
\end{align}
where for clarity all driving fields, messenger scales and mass parameters are
denoted by the same symbols $P$, $\La$ and $M$ respectively. It should be
noted that the driving fields only couple to one flavon each although all
possible combinations are permitted by charge conservation. This form can
always be achieved by a suitable rotation of the driving fields as described
in \cite{Antusch:2011sx}. Higher orders in $P$ are not relevant for the
alignment due to the vanishing vev of $P$.

All the flavon fields and their charges under shaping symmetries, as
well as their SU(5) and A$_5$ representations are listed
in Tab.~\ref{tab:flavonZn}. 
The messenger sector for the flavon alignment will be described
in appendix \ref{app:UV}.

\begin{table}
\centering
\begin{tabular}{l c l c c c c c c c c c c}
\toprule
& $\mathrm{SU(5)}$ & $\mathrm{A_5}$ & $\zz_{4}^R$ & $\zz_2$ & $\zz_2$ &
$\zz_3$ & $\zz_3$ & $\zz_3$ & $\zz_3$ & $\zz_3$ & $\zz_3$ & $\zz_4$ \\
\midrule
$\phi_2                             $ & $\mathbf{1}$ & $\mathbf{3 }$ & $0$ & $0$ & $0$ & $0$ & $0$ & $1$ & $2$ & $0$ & $0$ & $1$ \\
$\phi_3                             $ & $\mathbf{1}$ & $\mathbf{3 }$ & $0$ & $1$ & $1$ & $0$ & $2$ & $0$ & $2$ & $2$ & $0$ & $1$ \\
$\widetilde{\omega}                 $ & $\mathbf{1}$ & $\mathbf{5 }$ & $0$ & $0$ & $0$ & $1$ & $2$ & $1$ & $2$ & $0$ & $0$ & $1$ \\
$\omega                             $ & $\mathbf{1}$ & $\mathbf{5 }$ & $0$ & $0$ & $0$ & $0$ & $0$ & $0$ & $2$ & $0$ & $0$ & $0$ \\
$\lambda                            $ & $\mathbf{1}$ & $\mathbf{4 }$ & $0$ & $1$ & $0$ & $1$ & $1$ & $2$ & $2$ & $0$ & $0$ & $1$ \\
\midrule
$\theta_1                           $ & $\mathbf{1}$ & $\mathbf{1 }$ & $0$ & $1$ & $1$ & $0$ & $2$ & $2$ & $1$ & $1$ & $0$ & $0$ \\
$\theta_2                           $ & $\mathbf{1}$ & $\mathbf{1 }$ & $0$ & $1$ & $1$ & $0$ & $2$ & $1$ & $2$ & $1$ & $0$ & $3$ \\
$\theta_3                           $ & $\mathbf{1}$ & $\mathbf{1 }$ & $0$ & $0$ & $1$ & $0$ & $0$ & $1$ & $0$ & $1$ & $1$ & $3$ \\
$\epsilon_1                         $ & $\mathbf{1}$ & $\mathbf{1 }$ & $0$ & $0$ & $0$ & $0$ & $1$ & $1$ & $1$ & $0$ & $0$ & $0$ \\
$\epsilon_2                         $ & $\mathbf{1}$ & $\mathbf{1 }$ & $0$ & $0$ & $0$ & $0$ & $2$ & $0$ & $0$ & $0$ & $0$ & $3$ \\
$\epsilon_3                         $ & $\mathbf{1}$ & $\mathbf{1 }$ & $0$ & $1$ & $0$ & $1$ & $2$ & $0$ & $0$ & $0$ & $0$ & $0$ \\
$\epsilon_4                         $ & $\mathbf{1}$ & $\mathbf{1 }$ & $0$ & $0$ & $0$ & $2$ & $2$ & $2$ & $2$ & $0$ & $0$ & $3$ \\
$\epsilon_5                         $ & $\mathbf{1}$ & $\mathbf{1 }$ & $0$ & $1$ & $0$ & $1$ & $0$ & $2$ & $2$ & $0$ & $0$ & $3$ \\
\midrule
$D_\phi^{(2)}                       $ & $\mathbf{1}$ & $\mathbf{1 }$ & $2$ & $0$ & $0$ & $0$ & $0$ & $1$ & $2$ & $0$ & $0$ & $2$ \\
$D_\phi^{(3)}                       $ & $\mathbf{1}$ & $\mathbf{1 }$ & $2$ & $0$ & $0$ & $0$ & $2$ & $0$ & $2$ & $2$ & $0$ & $2$ \\
$D_{\widetilde{\omega}}             $ & $\mathbf{1}$ & $\mathbf{4 }$ & $2$ & $0$ & $0$ & $1$ & $2$ & $1$ & $2$ & $0$ & $0$ & $2$ \\
$D_{\omega \phi}^{(2)}              $ & $\mathbf{1}$ & $\mathbf{3'}$ & $2$ & $0$ & $0$ & $2$ & $1$ & $1$ & $2$ & $0$ & $0$ & $2$ \\
$D_{\omega \phi}^{(3)}              $ & $\mathbf{1}$ & $\mathbf{3'}$ & $2$ & $1$ & $1$ & $2$ & $2$ & $2$ & $2$ & $1$ & $0$ & $2$ \\
$D_\omega                           $ & $\mathbf{1}$ & $\mathbf{3'}$ & $2$ & $1$ & $0$ & $2$ & $2$ & $1$ & $2$ & $0$ & $0$ & $3$ \\
$D_\lambda                          $ & $\mathbf{1}$ & $\mathbf{5 }$ & $2$ & $0$ & $0$ & $1$ & $1$ & $2$ & $2$ & $0$ & $0$ & $2$ \\
$P                                  $ & $\mathbf{1}$ & $\mathbf{1 }$ & $2$ & $0$ & $0$ & $0$ & $0$ & $0$ & $0$ & $0$ & $0$ & $0$ \\
\bottomrule
\end{tabular}
\caption{The $\mathbb{Z}_n$ charges, $\mathrm{SU(5)}$ and $\mathrm{A_5}$
  representations of the flavons and driving fields.
  }
\label{tab:flavonZn}
\end{table}

\subsection{The Yukawa and mass matrices}
\label{sec:matter}

In this section we give the effective operators that determine the structure
of the Yukawa matrices and the right-handed neutrino mass matrix for the type I
seesaw \cite{seesaw} we implement. Note that the symmetries including shaping
symmetries are not sufficient to forbid all unwanted operators. Therefore
we have also studied a ``UV completion'' in appendix~\ref{app:UV} where
we give the renormalisable superpotential including messenger fields. After
integrating out the heavy vector-like messenger fields we end up with
the operators we are going to discuss in this section.

The matter content of our model is organised in ten-dimensional representations
of SU(5), $T_i$ with $i = 1$, 2, 3, five-dimensional representations $F$,
and one-dimensional representations $N$ which transform as one-, three- and
three-dimensional representations of A$_5$ respectively, see also Tab.~\ref{tab:matterZn}.

The superpotential for the neutrino sector reads
\begin{align}
\mathcal{W} &= y_1^n F N H_5 + y_2^n N N \omega \;.
\label{eq:superpot_neutrino}
\end{align}
After symmetry breaking this results in the Majorana
mass matrix 
\begin{align}
 M_{RR} &= y_2^n \begin{pmatrix}
2 \sqrt{\frac{2}{3}}(v_2 + v_3) & -\sqrt{3} v_2 & -\sqrt{3} v_2 \\
-\sqrt{3}v_2 & \sqrt{6}v_3 &- \sqrt{\frac{2}{3}}(v_2 + v_3)\\
-\sqrt{3}v_2 & -\sqrt{\frac{2}{3}}(v_2 + v_3) & \sqrt{6} v_3\\
\end{pmatrix} \;
\label{eq:massmatrix_righthanded}
\end{align}
for the right-handed neutrinos and the neutrino Yukawa matrix reads
in our basis
\begin{equation}
 Y_\nu = y_1^n \begin{pmatrix} 1 & 0 & 0 \\ 0 & 0 & 1 \\ 0 & 1 & 0  \end{pmatrix} \;.
 \label{eq:yukawamatrix_neutrino}
\end{equation}
Note that we are using the right-left convention for the Yukawa matrices,
which means that the first index of the matrix corresponds to the
SU(2)$_L$ doublets. Using the type I seesaw formula we end up
with the mass matrix for the light Majorana neutrinos

\begin{equation}
 m_{LL} = v_u^2 \frac{(y_1^{n})^2}{y_2^n} \begin{pmatrix}
a & b & b \\
b & c & d \\
b & d & c \\
\end{pmatrix}\;,
\end{equation}
where $v_u$ denotes the SU(2)$_L$ Higgs doublet vev of $H_5$
and the coefficients $a$, $b$, $c$, $d$ are functions of $v_2$ and $v_3$: 
\begin{align*}
a&\equiv -\frac{\sqrt{3/2}(v_{2} -2v_{3})}{4v_{3}^2 + 2v_{3} v_{2}-11v_{2}^2}~,\\
b&\equiv -\frac{3 \sqrt{3} v_{2}}{-8v_{3}^2 -4v_{3} v_{2}+22v_{2}^2}~, \\
c&\equiv \frac{3\sqrt{3/2}(4v_{3}^2 + 4 v_{3} v_{2} -3 v_{2}^2)}{x}~,\\
d&\equiv \frac{\sqrt{3/2}(4v_{3}^2 + 8v_{3} v_{2} + 13v_{2}^2)}{x}~,\\
x&\equiv 32v_{3}^3 + 24v_{2} v_{3}^2-84 v_{3} v_{2}^2 - 22 v_{2}^3 \;.
\end{align*}
The phenomenology of these structures will be discussed in the next section.

The effective superpotentials for the charged lepton and down-type quark
sector is
\begin{align}
\begin{split}
\mathcal{W}_{d,l} =&~ \frac{y_{33}}{\Lambda^{2}} T_3 (F \phi_2)_{\mathbf{1}} H_{24} \bar{H}_5
+ \frac{y_{22}}{\Lambda^{3}} T_2 (F
\phi_3)_{\mathbf{1}} \theta_1 \bar{H}_5 H_{24} + \frac{y_{21}}{\Lambda^{4}}
T_1 (F \phi_3)_{\mathbf{1}} \theta_3 H_{24}^2 \bar{H}_{5} \\
 &+ \frac{y_{12}}{\Lambda^{4}} T_2 \left(F \left( \phi_2 \phi_3 \right)_{\mathbf{3}}\right)_{\mathbf{1}} \theta_2 \bar{H}_5 H_{24}
+ \frac{y_{32}}{\Lambda^3} T_2 (F \phi_2)_{\mathbf{1}} \epsilon_1 \bar{H}_5 H_{24} \;,
\end{split} 
\label{eq:effectiveSuperpotential}
\end{align} 
where $\Lambda$ denotes a generic mass scale of the messenger fields (see appendix
\ref{app:UV} for more details). Note that the messenger sector plays a crucial
role here. Only by symmetries additional operators would be allowed and we would not
end up with the desired structures.

After plugging in the SU(5) and A$_5$ breaking vevs we find the following Yukawa
matrices for the down-type quarks
\begin{align}
Y_d &=  \begin{pmatrix}
0 &  \frac{1}{\Lambda^4}y_{12} v_{\phi}^ {(3)} v_{\phi}^{(2)} v_{\theta_2} & 0 \\
 \frac{y_{21}}{\Lambda^4} v_{\theta_3} v_{\phi}^ {(3)} & \frac{y_{22}}{\Lambda^3}v_{\phi}^ {(3)} v_{\theta_1} &
 0 \\
0 & \frac{y_{32}}{\Lambda^3} v_{\phi}^{(2)} v_{\epsilon_1} & \frac{y_{33}}{\Lambda^2} v_{\phi}^{(2)}\\
\end{pmatrix} \equiv \begin{pmatrix}
0 & a_{12} & 0\\
a_{21} & a_{22} & 0\\
0 & a_{32} & a_{33}\\
\end{pmatrix}\;,
\end{align}
and for the charged leptons
\begin{align}
Y_e &=  \begin{pmatrix}
0 & - 1/2 a_{21} & 0\\
6 a_{12} & 6 a_{22} & 6 a_{32}\\
0 & 0 & - 3/2 a_{33}\\
\end{pmatrix}~\; .
\label{eq:yuk_charged}
\end{align}
Note, first of all, that we find the SU(5) relation $Y_d = Y_e^T$ up
to order one CG coefficients. These coefficients
are arranged such that we have realistic Yukawa coupling ratios,
cf.~\cite{Antusch:2009gu, Antusch:2012fb, Antusch:2013rxa}, and we will as well be able
to correct the reactor mixing angle to realistic values.

In the up-type quark sector we have only used singlet flavons which
acquire a non-zero vev. The effective superpotential reads
\begin{align}
\begin{split}
\mathcal{W}_u &= \frac{y^u_{11}}{\Lambda^3}\epsilon_1 \epsilon_2 \epsilon_4 T_1 T_1 H_5
+ \frac{y_{12}^u}{\Lambda^3} T_1 T_2 H_5 \epsilon_1 \epsilon_2 \epsilon_3\\
 &+ \frac{y^u_{22}}{\Lambda^2} T_2 T_2 H_5 \epsilon_1 \epsilon_1 +
\frac{y_{31}^u}{\Lambda^2} T_1 T_3 H_5 \epsilon_1 \epsilon_5
+ \frac{y^u_{32}}{\Lambda} T_3 T_2 H_5 \epsilon_1 + y^u_{33} T_3 T_3 H_5 \;,
\end{split} 
\end{align}
and from that we find for the up-type quark Yukawa matrix
\begin{align}
Y_u &= \begin{pmatrix}
\frac{y_{11}^u}{\Lambda^2} v_{\eps_1} v_{\eps_2} v_{\eps_4} &
\frac{y_{12}^u}{\Lambda^2} v_{\eps_1} v_{\eps_2} v_{\eps_3} &
\frac{y_{31}^u}{\Lambda^2} v_{\eps_5} v_{\eps_1} \\
\frac{y_{12}^u}{\Lambda^2} v_{\eps_1} v_{\eps_2}v_{\eps_3} & \frac{y_{22}^u}{\Lambda^2} v_{\eps_1}^2 &
\frac{y_{32}^u}{\Lambda} v_{\eps_1}\\
\frac{y_{31}^u}{\Lambda^2} v_{\eps_5} v_{\eps_1} & \frac{y_{32}^u}{\Lambda} v_{\eps_1} & y_{33}^u 
\end{pmatrix} \equiv \begin{pmatrix}
b_{11} & b_{12} & b_{13} \\
b_{12} & b_{22} & b_{23} \\
b_{13} & b_{23} & b_{33} 
\end{pmatrix}~.
\end{align}

The complete matter and Higgs field content of the model and their charges
under additional shaping symmetries is collected in Tab.~\ref{tab:matterZn}.
We have checked that there are no new additional effective operators
contributing to the Yukawa matrices up to mass dimension eight. Hence,
we expect possible higher order corrections to be negligible small.
We will comment more on this in appendix~\ref{app:UV} where
we discuss the messenger sector of the model.

\begin{table}
\centering
\begin{tabular}{l c l c c c c c c c c c c}
\toprule
 & $\mathrm{SU(5)}$ & $\mathrm{A_5}$ & $\zz_{4}^R$ & $\zz_2$ & $\zz_2$ &
$\zz_3$ & $\zz_3$ & $\zz_3$ & $\zz_3$ & $\zz_3$ & $\zz_3$ & $\zz_4$ \\
\midrule
$F                                  $ & $\mathbf{\bar{5}} $ & $\mathbf{3}$ & $1$ & $0$ & $0$ & $0$ & $0$ & $1$ & $2$ & $0$ & $0$ & $0$ \\
$N                                  $ & $\mathbf{1}       $ & $\mathbf{3}$ & $1$ & $0$ & $0$ & $0$ & $0$ & $0$ & $2$ & $0$ & $0$ & $2$ \\
$T_1                                $ & $\mathbf{10}      $ & $\mathbf{1}$ & $1$ & $1$ & $0$ & $2$ & $2$ & $2$ & $2$ & $0$ & $0$ & $0$ \\
$T_2                                $ & $\mathbf{10}      $ & $\mathbf{1}$ & $1$ & $0$ & $0$ & $0$ & $2$ & $1$ & $1$ & $0$ & $0$ & $3$ \\
$T_3                                $ & $\mathbf{10}      $ & $\mathbf{1}$ & $1$ & $0$ & $0$ & $0$ & $0$ & $2$ & $2$ & $0$ & $0$ & $3$ \\
\midrule
$H_5                                $ & $\mathbf{5}       $ & $\mathbf{1}$ & $0$ & $0$ & $0$ & $0$ & $0$ & $2$ & $2$ & $0$ & $0$ & $2$ \\
$\bar{H}_5                          $ & $\mathbf{\bar{5}} $ & $\mathbf{1}$ & $0$ & $0$ & $0$ & $2$ & $1$ & $2$ & $0$ & $0$ & $1$ & $0$ \\
$H_{24}                             $ & $\mathbf{24}      $ & $\mathbf{1}$ & $0$ & $0$ & $0$ & $1$ & $2$ & $0$ & $0$ & $0$ & $2$ & $0$ \\
$H'_{24}                       $ & $\mathbf{24      }$ & $\mathbf{1}$ & $2$ &
$0$ & $0$ & $2$ & $1$ & $0$ & $0$ & $0$ & $1$ & $0$ \\
$S                                  $ & $\mathbf{1}$ & $\mathbf{1 }$ & $2$ & $0$ & $0$ & $1$ & $2$ & $0$ & $0$ & $0$ & $2$ & $0$ \\
\bottomrule
\end{tabular}
\caption{Charges under $\mathbb{Z}_n$ and $\mathrm{SU(5)}$ and
  $\mathrm{A_5}$ representations of the matter and Higgs fields.}
\label{tab:matterZn}
\end{table}

\section{Phenomenology}
\label{sec:Phenomenology}

In this section we present the phenomenological implications of our model. First we discuss
the quarks and charged leptons. We put a special emphasis on the Yukawa coupling ratios of
the charged leptons and down-type quarks which arise in our model. Afterwards we discuss briefly
a numerical fit to the low energy charged lepton and quark masses and CKM mixing parameters.
In the second part of this section we cover the neutrino sector of our model. We revise the neutrino
mass sum rule and show how corrections for the leptonic mixing parameters occur due to a
non-diagonal charged lepton Yukawa matrix and RGE corrections. Finally, we show the predictions of
our model for the leptonic mixing parameters and  for observables testable in the near future in
neutrino experiments.

\subsection{The quark and charged lepton sector}

In the last section we derived the Yukawa matrices for the quark and the charged lepton sector
which fulfill the minimal SU(5) relation $Y_{d}=Y_{e}^{\text{T}}$ up to $\mathcal{O}(1)$ CG
coefficients. This deviation from the minimal relation gives better agreement to the observed
fermion masses \cite{Antusch:2009gu, Antusch:2012fb, Antusch:2013rxa}. To be concrete, we have the ratios
\begin{align}
\frac{y_{e}}{y_{d}}   \approx\frac{1}{2} \;, \quad \frac{y_{\mu}}{y_{s}} \approx 6 \;, \quad
\frac{y_{\tau}}{y_{b}} \approx\frac{3}{2} \;,
\end{align}
where $y_{\tau},~y_{\mu},~y_{e},~y_{b},~y_{s}$ and $y_{d}$ are the eigenvalues of the
Yukawa matrices $Y_{e}$ and $Y_{d}$. Especially, the relation for the third generation
was already realised to be very promising in \cite{Antusch:2009gu} and then its phenomenology
was further studied in subsequent publications, e.g.~\cite{Antusch:2011sq, Antusch:2011xz, Antusch:2012gv}.

In \cite{Antusch:2013jca} the double ratio
\begin{equation}
\frac{y_{\mu}}{y_{s}}\frac{y_{d}}{y_{e}}\approx 10.7^{+1.8}_{-0.8}
\end{equation}
was studied which depends only weakly on RGE corrections and
supersymmetric threshold corrections. Plugging in our results for the Yukawa
coupling ratios we get $\frac{y_{\mu}}{y_{s}}\frac{y_{d}}{y_{e}}=12$ which is
within 1$\sigma$ as was already realised in \cite{Antusch:2013jca}.
In contrast, the very popular Georgi-Jarlskog relations \cite{Georgi:1979df},
$y_{\mu}/y_{s}=3$ and  $y_{e}/y_{d}=1/3$, yield $\frac{y_{\mu}}{y_{s}}\frac{y_{d}}{y_{e}}=9$
which deviates more than 2$\sigma$ from the best fit result.
 
Since we use right-left convention we have to diagonalise $Y_e$ via
$U_{e}^{\dagger}Y_{e}^{\dagger}Y_{e}U_{e}=\text{Diag}(y_{e}^{2},y_{\mu}^{2},y_{\tau}^{2})$
where $U_{e}=U_{12}U_{13}U_{23}$ is a unitary matrix. $U_{23},U_{13}$ and $U_{12}$ are given as
\begin{equation}
U_{23}=
\begin{pmatrix}
1&0&0\\
0&c_{23}^{e}&s^{e}_{23}\text{e}^{-\text{i }\delta_{23}^{e}}\\
0&-s_{23}^{e}\text{e}^{\text{i }\delta_{23}^{e}}&c^{e}_{23}\\
\end{pmatrix}
\label{eq:U23}
\end{equation}
and analogous expressions for $U_{12}$ and $U_{13}$.
We use the abbreviation $\cos(\theta_{ij}^{e})=c_{ij}^{e}$ and
$\sin(\theta_{ij}^{e})=s_{ij}^{e}$.
Bearing in mind that $\theta_{13}^{e} = \theta_{23}^{e} = 0$
in a very good approximation, the matrix $U_{e}$ is
parameterised only by one angle $\theta_{12}^{e}$ and one phase
$\delta_{12}^{e}$.

If we compare both sides of $U_{e}^{\dagger}Y_{e}^{\dagger}Y_{e}U_{e}=\text{Diag}(y_{e}^{2},y_{\mu}^{2},y_{\tau}^{2})$
we find at leading order
\begin{equation}
\theta_{12}^{e} = \left| \frac{a_{12}}{a_{22}} \right| \text{ and } \delta_{12}^{e} = \arg \frac{a_{12}}{a_{22}} \;.
\label{eq:theta_122}
\end{equation}
The eigenvalues of $Y_e$ and $Y_d$ are not sufficient to fix the values of $a_{12}$ and $a_{21}$
independently since at leading order only their product appears in the expression for the eigenvalues.
And importantly, the phase $\delta_{12}^{e}$ is essentially undetermined by the quark and charged lepton
sector only.
Nevertheless, neglecting mixing from the up-type quark sector the same procedure for the down-type sector
leads to the relation $\theta_{C} \approx |\tfrac{a_{21}}{a_{22}}|$ for the Cabibbo angle. And in this
case it follows for $\theta_{12}^{e}$ \cite{Antusch:2012fb}
\begin{equation}
\theta_{12}^{e} \approx \theta_{C}  \;,
\label{eq:theta_C1}
\end{equation}
and subsequently $\theta_{13}^{\text{PMNS}} \approx \theta_C/\sqrt{2}$
\cite{Marzocca:2011dh,Antusch:2011qg, Antusch:2012fb}.

The main focus of this paper lies on the neutrino sector and for that especially $y_\tau$ and $\theta_{12}^e$
are important. To quantify them we have fitted the parameters of the Yukawa matrices at the high energy scale
to the low energy observables with the help of the REAP package \cite{Antusch:2005gp}. The Yukawa coupling ratios
we discussed before are only valid in a regime with rather large $\tan \beta \approx 30$ where we have to consider
so-called SUSY threshold corrections for the masses and mixing parameters \cite{SUSYthresholds}.

The approach we have used here is documented, for instance, in \cite{Antusch:2008tf, Spinrath:2010dh, Antusch:2011sq}
so that we will not go into much detail here. For the up-type quarks we have used the tree-level MSSM matching relation
\begin{equation}
Y_u^{\text{MSSM}} = \frac{Y_{u}^{\text{SM}} }{\sin\beta} \;
\label{eq:y_mssm_u}
\end{equation}
at the SUSY scale $M_{\text{SUSY}} = 1$~TeV.
For the Yukawa couplings of the charged leptons and down-type quarks we have included the $\tan \beta$
enhanced threshold corrections in the matching formulas
\begin{align}
y^{\text{MSSM}}_{e,\mu,\tau}&=\frac{y_{e,\mu,\tau}^{\text{SM}}}{\cos\beta(1+\epsilon_{l}\tan\beta)} \;,\\
y^{\text{MSSM}}_{d,s}&=\frac{y_{d,s}^{\text{SM}}}{\cos\beta(1+\epsilon_{q}\tan\beta)} \;,\\
y^{\text{MSSM}}_{b}&=\frac{y_{b}^{\text{SM}}}{\cos\beta(1+(\epsilon_{q}+\epsilon_{A})\tan\beta)} \;.
\end{align}
Also the quark mixing parameters are modified by this matching via
\begin{align}
\theta_{i3}^{\text{MSSM}}&=\frac{\theta_{i3}^{\text{SM}}(1+(\epsilon_{q}+\epsilon_{A})\tan\beta)}{1+\epsilon_{q}\tan\beta} \;,\\
\theta_{12}^{\text{MSSM}}&=\theta_{12}^{\text{SM}} \;,\\
\delta^{\text{MSSM}}_{\text{CKM}}&=\delta^{\text{SM}}_{\text{CKM}} \;.
\end{align}

Hence, apart from the parameters in the Yukawa matrices we have two additional parameters
to describe the SUSY threshold corrections. For definiteness we have fixed $\tan \beta = 30$
and $M_{\text{GUT}} = 2\cdot 10^{16}$~GeV.

\begin{table}
\centering
\begin{tabular}{c c }
\toprule
Parameter& Value \\
\midrule
	$a_{12}$&$4.46\cdot 10^{-4}$\\
	$a_{22}$&$2.12\cdot 10^{-3}$\\
	$a_{21}$&$5.95\cdot 10^{-4}$\\
	$a_{32}$&$-1.22\cdot 10^{-3}$\\
	$a_{33}$&$1.5\cdot 10^{-1}$\\
	\midrule
	$b_{11}$&$-2.22\cdot 10^{-7}$\\
	$b_{12}$&$9.54\cdot 10^{-5}$\\
	$b_{13}$&$1.19\cdot 10^{-3}$\\
	$b_{22}$&$1.72\cdot 10^{-3}$\\
	$b_{23}$&$1.29\cdot 10^{-2}$\\
	$b_{33}$&$5.19\cdot 10^{-1}$\\
	$\delta_{12}^{u}$&$5.78$\\
	$\delta_{13}^{u}$&$6.16\cdot 10^{-1}$\\
	$\delta_{23}^{u}$&0\\
	\midrule
	$\epsilon_{q}\tan \beta$&$0.36$\\
	$\epsilon_{A}\tan \beta$&$0.19$\\
	\bottomrule
\end{tabular}
\caption{
Parameters of the quark and charged leptons Yukawa matrices
at the GUT scale with $\tan \beta=30$ and $M_{\text{SUSY}}=1$~TeV.
}
\label{tab:yuk_parameter}
\end{table}

\begin{table}
 \centering
 
 \begin{tabular}{c c}
\toprule
Quantity (at $m_t(m_t)$)& Experiment  \\ \midrule
$y_\tau$ in $10^{-2}$   & 1.00   \\
$y_\mu$ in $10^{-4}$    & 5.89   \\
$y_e$ in $10^{-6}$  & 2.79  \\
\midrule
$y_b$ in $10^{-2}$      & $1.58 \pm 0.05$ \\
$y_s$ in $10^{-4}$  & $2.99 \pm 0.86$ \\
$y_s/y_d$           & $18.9 \pm 0.8$  \\
\midrule
$y_t$               & $0.936 \pm 0.016$ \\
$y_c$ in $10^{-3}$      & $3.39 \pm 0.46$ \\
$y_u$ in $10^{-6}$      & $7.01^{+2.76}_{-2.30}$ \\
\midrule
$\theta_{12}^{\text{CKM}}$ & $0.2257^{+0.0009}_{-0.0010}$  \\[0.3pc]
$\theta_{23}^{\text{CKM}}$ & $0.0415^{+0.0011}_{-0.0012}$  \\[0.1pc]
$\theta_{13}^{\text{CKM}}$ & $0.0036 \pm 0.0002$           \\[0.1pc]
$\delta_{\text{CKM}}$ & $1.2023^{+0.0786}_{-0.0431}$       \\
\bottomrule
\end{tabular}
\caption{
Experimental data for the quark and charged leptons Yukawa couplings at low
energy taken from \cite{Xing:2007fb} and the mixing angles were taken from
\cite{Beringer:1900zz}. The uncertainties for the charged lepton Yukawa couplings
were assumed to be $1\%$, for more details see the text. Our fit to these
observables has $\chi^2 \approx 0.05$.
}
\label{tab:experiment}
\end{table}

We performed a $\chi^{2}$-fit to the low energy observables (nine fermion  masses,
three mixing angles, one phase). Since we have more parameters than observables it is
not surprising that we find $\chi^2 \approx 0.05$ where we stopped the time consuming
minimisation procedure because the fit is sufficiently good. Note, that in principle
 $\chi^2$ can be made arbitrarily small.
The numerical results for the parameters can be found in Tab.~\ref{tab:yuk_parameter}.
For convenience we have also collected the low energy observables including their uncertainties
in Tab.~\ref{tab:experiment}. Note, that we have assumed an uncertainty of 1$\%$ of
the Yukawa couplings for the charged leptons which is larger than their experimental errors. But
since we use only one-loop RGEs we cannot expect a very high precision.

\subsection{Neutrino sector}

In this section we present the phenomenological implications for the neutrino
sector of our model. First, we revise the mass sum rule present in our model
which was also discussed before in other golden ratio models with an
$\text{A}_{5}$ family symmetry \cite{Everett:2008et,Cooper:2012bd}.
Then we discuss two important corrections in our model. First we study
RGE corrections and then corrections from the charged lepton sector
to the neutrino mixing angles and phases in terms of sum rules.
Especially, the latter is crucial to predict the reactor mixing angle,
within its experimentally allowed range.

Including RGE effects
rules out the inverted neutrino mass hierarchy in our setup
because of incompatible constraints from the mass and the mixing sum
rule on the one hand and the experimental value for $\theta_{12}^{\text{PMNS}}$
on the other hand.

Finally, we will discuss the results from a numerical parameter scan
for various observables in the neutrino sector.

\subsubsection{The neutrino mass sum rule}

The neutrino sector is described  by the superpotential from
eq.~\eqref{eq:superpot_neutrino}. The right-handed neutrino mass matrix
in eq.~\eqref{eq:massmatrix_righthanded} is diagonalised by the golden
ratio mixing matrix $U_{\text{GR}}$ from eq.~\eqref{eq:U_GR}
\begin{equation}
U_{\text{GR}}^{\text{T}}M_{\text{RR}}U_{\text{GR}}=\text{Diag}(M_{1}, M_{2},M_{3})
\end{equation}
with the heavy neutrino masses
\begin{align}
M_{1}&=\frac{y_{2}(v_{2}(6\phi_{g}-2)+4v_{3})}{\sqrt{6}} \;,\\
M_{2}&=\frac{y_{2}(4v_{3}-v_{2}(\frac{6}{\phi_{g}}+2))}{\sqrt{6}} \;,\\
M_{3}&=\frac{y_{2}\sqrt{2}(v_{2}+4v_{3})}{\sqrt{3}} \;.
\end{align}
These masses obey the sum rule
\begin{equation}
M_{1}+M_{2}=M_{3} \;,
\label{eq:sumrule_schwer}
\end{equation}
which was already noted in \cite{Cooper:2012bd}.

The light neutrino mass matrix $m_{LL}$ in eq.~\eqref{eq:yukawamatrix_neutrino} 
is as well diagonalised by $U_{\text{GR}}$ after a matrix $P^{\prime}=\text{Diag}(1,1,-1)$ with
unphysical phases has been applied to $U_{\text{GR}}$ \cite{Cooper:2012bd}.

The resulting complex light neutrino masses $m_{i}$ read
\begin{align}
m_{1}&=\frac{\sqrt{6}y^{2}v_{u}^{2}}{y_{2}(v_{2}(6\phi_{g}-2)+4v_{3})} \;,\\
m_{2}&=\frac{\sqrt{6}v_{u}^{2}y^{2}}{y_{2}(4v_{3}-v_{2}(\frac{6}{\phi_{g}}+2))} \;,\\
m_{3}&=\frac{\sqrt{\frac{3}{2}}y^{2}v_{u}^{2}}{(v_{2}+4v_{3})y_{2}} \;
\end{align}
which obey the inverse sum rule \cite{Ding:2011cm,Cooper:2012bd}
\begin{equation}
\frac{1}{m_{1}}+\frac{1}{m_{2}}=\frac{1}{m_{3}} \;.
\label{eq:sumrules_neutrino}
\end{equation}

In this sum rule the neutrino masses are still complex.
If we want to discuss the physical masses we have to consider the absolute values of
the masses $\left|m_{i}\right|$. We reexpress the mass $m_{i}$ as
$m_{i}=\left|m_{i}\right| \text{exp}(-\text{i }\alpha_{i})$. One phase $\alpha_{i}$
is unphysical since it corresponds to a global phase of the neutrino mass matrix.
We choose the mass $m_{3}$ to be real and set $\alpha_{3} = 0$. The phases $\alpha_1$
and $\alpha_2$ are then the Majorana phases. 

Writing down the Majorana phases explicitly the sum rule from eq.~\eqref{eq:sumrules_neutrino}
reads
\begin{equation}
\frac{\text{e}^{\text{i} \, \alpha_{1}}}{\left|m_{1}\right|}+\frac{\text{e}^{\text{i} \, \alpha_{2}}}{\left|m_{2}\right|}=\frac{1}{\left|m_{3}\right|} \;.
\label{eq:masses_sum}
\end{equation}
One can rewrite the sum rule using the mass squared differences which yields a
mass range for the lightest neutrino mass in both hierarchies \cite{Barry:2010yk},
see also \cite{King:2013psa}.
But note that this sum rule is valid at the seesaw scale and hence the mass sum rule
should be evaluated at this high scale.

\subsubsection{Renormalisation Group Corrections}
\label{sec:RGE}

Since the experimental values for the mixing angles and the mass squared differences
were measured at a low energy scale in contrast to the model parameters which are defined
at a high energy scale, possible effects due to RGE corrections have to be considered.

The RGE corrections for the mass squared differences were derived, for instance, in \cite{Antusch:2003kp}
\begin{align}
8\pi^{2}\frac{\text{d}}{\text{d}t}\Delta m^{2}_{21}&=\alpha \Delta m^{2}_{21}+Cy_{\tau}^{2}\left[2 s_{23}^{2}\left(m_{2}^{2}c_{12}^{2}-m_{1}^{2}s_{12}^{2}\right)+F_{\text{sol}}\right] \;,\\
8\pi^{2}\frac{\text{d}}{\text{d}t}\Delta m^{2}_{32}&=\alpha \Delta m^{2}_{32}+Cy_{\tau}^{2}\left[2 c_{23}^{2} m^{2}_{3}c_{13}^{2}-2 m_{2}^{2}c_{12}^{2}s_{23}^{2}+F_{\text{atm}}\right] \;,
\label{eq:rge_masses}
\end{align}
where 
\begin{align}
F_{\text{sol}}&=\left(m_{1}^{2}+m_{2}^{2}\right) s_{13} \sin 2\th_{12}^{\text{PMNS}} \sin 2\th_{23}^{\text{PMNS}} \cos \delta_{\text{PMNS}}\\
 &+\nonumber 2 s_{13}^{2} c_{23}^{2}\left(m_{2}^{2} s_{12}^{2}-m_{1}^{2} c_{12}^{2}\right),\\
F_{\text{atm}}&=-m_{2}^{2}s_{13} \sin 2\th_{12}^{\text{PMNS}} \sin 2\th_{23}^{\text{PMNS}} \cos \delta_{\text{PMNS}} -2 m_{2}^{2}s_{13}^{2}s_{12}^{2} c_{23}^{2}
\end{align}
and $t= \ln \mu$.
In our analytical estimates we will neglect $F_{\text{sol}}$ and  $F_{\text{atm}}$
because they are proportional to the small $s_{13}$. The term proportional to $\alpha \approx 1/137$
is negligible as well.
If we also neglect the running of the parameters in the $\beta$ functions we
can integrate the RGEs and obtain approximations for the mass squared differences at
the seesaw scale $M_S \approx 10^{13}$~GeV using the best-fit values for the observables.
Together with the mass sum rule this implies an allowed range for the neutrino mass scale
\begin{align}
  0.011 \text{ eV} &\lesssim  m_1 \phantom{\lesssim 0.454 \text{ eV}} \text{  for NH,} \label{eq:NHMassRange} \\
  0.028 \text{ eV} &\lesssim  m_3 \lesssim 0.454 \text{ eV} \text{ for IH.} \label{eq:IHMassRange}
\end{align}
Note that the sum rule only implies a lower bound on the mass scale for the normal hierarchy.

The analytical RGE expressions for the mixing angles of the PMNS matrix are \cite{Antusch:2003kp}
\begin{align}
\dot{\theta}_{12}^{\text{PMNS}}&=-\frac{C y_{\tau}^{2}}{32\pi^{2}}\sin 2\theta_{12}^{\text{PMNS}} s^{2}_{23}\frac{\left|m_{1}\text{e}^{\text{i}\alpha_{1}}+m_{2}\text{e}^{\text{i}\alpha_{2}}\right|^{2}}{\Delta m_{21}^{2}}+\mathcal O(\theta_{13}^{\text{PMNS}}) \;,
\label{eq:theta_12_r} \\
\dot{\theta}_{13}^{\text{PMNS}}&=\frac{C y_{\tau}^{2}}{32\pi^{2}}\sin 2\theta_{12}^{\text{PMNS}} \sin 2\theta_{23}^{\text{PMNS}} \frac{m_{3}}{\Delta m_{32}^{2}(1+\zeta)}\\&\nonumber \times
\left[m_{1}\cos(\alpha_{1}-\delta_{\text{PMNS}})-(1+\zeta)m_{2}\cos(\alpha_{2}-\delta_{\text{PMNS}})-\zeta m_{3}\cos\delta_{\text{PMNS}} \right]+\mathcal O(\theta_{13}^{\text{PMNS}}) \;, \label{eq:theta_13_r} \\
\dot{\theta}_{23}^{\text{PMNS}}&=-\frac{C y_{\tau}^{2}}{32\pi^{2}}\sin 2\theta_{23}^{\text{PMNS}} \frac{1}{\Delta m_{32}^{2}}\left[c_{12}^{2}\left|m_{2}\text{e}^{\text{i}\alpha_{2}}+m_{3}\right|^{2}+s_{12}^{2}\frac{\left|m_{1}\text{e}^{\text{i}\alpha_{1}}+m_{3}\right|^{2}}{1+\zeta}\right]\\&+\nonumber\mathcal O(\theta_{13}^{\text{PMNS}}) \;. \label{eq:theta23_running}
\end{align}
Here the abbreviation $\zeta=\frac{\Delta m_{21}^{2}}{\Delta m_{32}^{2}}$ was used. In the MSSM $C=1$
and $\frac{C y_{\tau}^{2}}{32\pi^{2}}\approx 0.3\cdot 10^{-6}(1+\tan^{2}\beta)$, where we set
$\tan \beta = 30$.

The running of $\theta_{12}^{\text{PMNS}}$ can be enhanced by the small mass squared difference in the
denominator if the mass scale is much larger than the splitting. Hence, for heavy masses all mixing angles can
change considerably. This will be especially important for the inverted hierarchy.

Before we will come back to this we just want to give here the value
for $\theta_{12}^{\text{PMNS}}$ at $M_S$ depending on the mass scale. In order to determine the value of
$\theta_{12}^{\text{PMNS}}(M_S)$ we need to calculate the difference of the Majorana phases
$\Delta=\alpha_1-\alpha_2$ at the seesaw scale. The absolute value of the mass
sum rule,
cf.~eq.~\eqref{eq:sumrules_neutrino}, implies

\begin{align}
\cos\Delta &= \frac{1}{2} m_1 m_2\left(\frac{1}{m_3^2}-\frac{1}{m_2^2}-\frac{1}{m_1^2}\right) \;,
\end{align}
where the masses label here the absolute values of the neutrino masses.
Inserting this expression as well as the mass squared differences at the high scale leads to
\begin{align}
 \theta_{12}^{\text{PMNS}} (M_S) &\approx \left( 23.00 - 2170.02 \frac{m_3^2}{\text{ eV}^2} - \frac{0.013}{m_3^2} \text{eV}^2 \right)^\circ \text{ for IH.}
\end{align}
The same expression for the normal hierarchy is rather lengthy and not that relevant for our discussion
so that we do not quote it explicitly here.
With the minimal value of $m_{3}$ from eq.~\eqref{eq:IHMassRange} we find the maximal value for $\theta_{12}^{\text{PMNS}} (M_S)$
\begin{align}
\theta_{12}^{\text{PMNS}} (M_S)\approx 5.65^{\circ} \text{ for IH.} 
\label{eq:IHth12maxvalue}
\end{align}
Performing the same analysis for the normal hierarchy with the minimal value for $m_1$, cf.\ eq.~\eqref{eq:NHMassRange}, yields
\begin{align}
\theta_{12}^{\text{PMNS}} (M_S)\approx 33.44^{\circ}\text{ for NH.} 
\label{eq:NHth12maxvalue}
\end{align}
As we can see for the inverted hierarchy case we find
an inevitable sizeable running for $\tan \beta = 30$. 

\subsubsection{Corrections from the charged lepton sector}
\label{sec:Ye}

If we assume a non-diagonal Yukawa matrix of the charged leptons, their mixing angles
influence the parameters of the PMNS matrix via $U_{\text{PMNS}}=U_{e}^{\dagger} U_{\nu}$
with the neutrino mixing matrix $U_{\nu}$ and the mixing matrix of the charged leptons $U_{e}$. 
As we discussed before in our model $U_{\nu}$ is of the golden ratio form $U_{\text{GR}}$, cf.\
eq.~\eqref{eq:U_GR}.

Approximate expressions for the leptonic mixing angles in terms of sum rules of neutrino mixing
angles and the charged lepton mixing angles were derived, for instance, in \cite{King:2002nf,Antusch:2005kw,Antusch:2008yc}.
In leading order in the small mixing angles they read
\begin{align}
\label{eq:sum_rules1}
s_{23}^{\text{PMNS}} \text{e}^{-\text{i} \delta_{23}} &\approx
s_{23}^{\nu} \text{e}^{-\text{i}\delta_{23}^{\nu}}-\theta_{23}^{e} c_{23}^{\nu} \text{e}^{-\text{i}\delta_{23}^{e}} \;,\\
\label{eq:sum_rules2}
\theta_{13}^{\text{PMNS}}\text{e}^{-\text{i} \delta_{13}} &\approx \theta_{13}^{\nu} \text{e}^{-\text{i}\delta_{13}^{\nu}}-\theta_{13}^{e} c_{23}^{\nu} \text{e}^{-\text{i}\delta_{13}^{e}}-\theta_{12}^{e} s_{23}^{\nu} \text{e}^{\text{i}(-\delta_{23}^{\nu}-\delta_{12}^{e})} \;, \\
\label{eq:sum_rules3}
s_{12}^{\text{PMNS}}\text{e}^{-\text{i} \delta_{12}} &\approx s_{12}^{\nu} \text{e}^{-\text{i}\delta_{12}^{\nu}}+\theta_{13}^{e} c_{12}^{\nu} s_{23}^{\nu}\text{e}^{\text{i}(\delta_{23}^{\nu}-\delta_{13}^{e})}-\theta_{12}^{e}c_{23}^{\nu}c_{12}^{\nu}\text{e}^{-\text{i}\delta_{12}^{e}} \;.
\end{align}

In our model we have arranged $\theta_{12}^e \approx \theta_C$ and
$\theta_{13}^{e} \approx \theta_{23}^{e} \approx 0$.
This can easily be seen in the Yukawa matrix $Y_{e}$ from eq.~\eqref{eq:yuk_charged}
where the mixing between the generations is governed to leading order by the ratios of the
elements in the rows.

Using these estimates as well as the golden ratio mixing angles of the A$_5$ model
from eqs.~(\ref{eq:A5_Winkel1}, \ref{eq:A5_Winkel2}, \ref{eq:A5_Winkel3}) the expressions from eqs.~(\ref{eq:sum_rules1},~\ref{eq:sum_rules2},~\ref{eq:sum_rules3}) simplify to
\begin{align}
s_{23}^{\text{PMNS}}\text{e}^{-\text{i} \delta_{23}} &\approx  \frac{1}{\sqrt{2}}\text{e}^{-\text{i}\delta_{23}^{\nu}}+\theta_{23}^{\text{rad}} \;,
\label{eq:sum_rules1e} \\
\theta_{13}^{\text{PMNS}}\text{e}^{-\text{i} \delta_{13}} &\approx -\frac{1}{\sqrt{2}}\theta_{12}^{e} \text{e}^{\text{i}(-\delta_{23}^{\nu}-\delta_{12}^{e})}+\theta_{13}^{\text{rad}} \;,
\label{eq:sum_rules2e} \\
s_{12}^{\text{PMNS}}\text{e}^{-\text{i} \delta_{12}} &\approx s_{12}^{\nu} \text{e}^{-\text{i}\delta_{12}^{\nu}}-\frac{1}{\sqrt{2}}\theta_{12}^{e}c_{12}^{\nu}\text{e}^{-\text{i}\delta_{12}^{e}}+\theta_{12}^{\text{rad}} \;,
\label{eq:sum_rules3e}
\end{align}
where the extra terms $\theta^{\text{rad}}_{ij}$ are complex numbers
representing the RGE corrections.

It follows from eq.~\eqref{eq:sum_rules2e} that $\theta_{13}^{\text{PMNS}}$ is dominated by $\theta_{12}^{e}$
as long as the RGE corrections are not very large which leads to the already mentioned relation
$\theta_{13}^{\text{PMNS}} \approx \theta_{C}/\sqrt{2}$.

At the seesaw scale we can neglect the radiative corrections and
find the sum rule \cite{Antusch:2005kw}
\begin{equation}
 \theta_{12}^{\text{PMNS}} + \frac{1}{\sqrt{2}}\theta_{12}^e\cos\left(\delta_{\text{PMNS}}-\pi\right)\approx \theta_{12}^{\nu}.
 \label{eq:allow_range_th12MS}
\end{equation}
Since $\theta_{13}^{\text{PMNS}} \approx \theta_{C}/\sqrt{2}$ the possible
values for $\theta_{12}^{\text{PMNS}}$ at the seesaw scale are hence
restricted to be in the range $(24 - 39)^\circ$.

\subsubsection{Results for inverted hierarchy}

Using the previous results it is easy to understand that for the inverted
hierarchy we do not find any allowed parameter points for $\tan \beta = 30$.
As we have discussed before the allowed range for $\theta_{12}^{\text{PMNS}}$ at the
seesaw scale is $(24 - 39)^\circ$ cf.~eq.~\eqref{eq:allow_range_th12MS}. On the other
hand from  eq.~\eqref{eq:IHth12maxvalue} we find
$\theta_{12}^{\text{PMNS}}$ at the seesaw scale to be smaller than 5.65$^\circ$
and hence the inverted hierarchy is not viable.

In this way we can also estimate that the inverted hierarchy is only possible
in this setup for $\tan \beta \lesssim 17$ to keep the RGE corrections small enough
which is nevertheless in tension with our Yukawa coupling ratios,
cf.~\cite{Antusch:2009gu}.

Note that the RGE running is quite sizeable and hence our approximations might
not be justified. But even in a numerical scan using the REAP package \cite{Antusch:2005gp}
we did not find any viable points which we can understand at least qualitatively
from our estimates.

\subsubsection{Results for normal hierarchy}

In our analytical estimates we find an overlap for the allowed ranges for $\theta_{12}^{\text{PMNS}}$,
cf.\ eq.~\eqref{eq:NHth12maxvalue} and \eqref{eq:allow_range_th12MS}, and hence the normal hierarchy
is feasible here.

We find an allowed parameter space which is compatible
within 3$\sigma$ with all observables. In our setup the neutrino sector is completely
determined by four parameters. Two real parameters and one phase in the effective
light neutrino mass matrix and one additional phase from the charged lepton sector
($\delta_{12}^e$). Note, that $\theta_{12}^e$
was already fixed in the fit and we will find that $\theta_{13}^{\text{PMNS}}$ is in
the correct range.
For our parameter scan we have used again the REAP package \cite{Antusch:2005gp}, where we have
set the seesaw scale to about $10^{13}$~GeV and $y_1^n = 0.1$.

\begin{figure}
\centering
\includegraphics[scale=0.49]{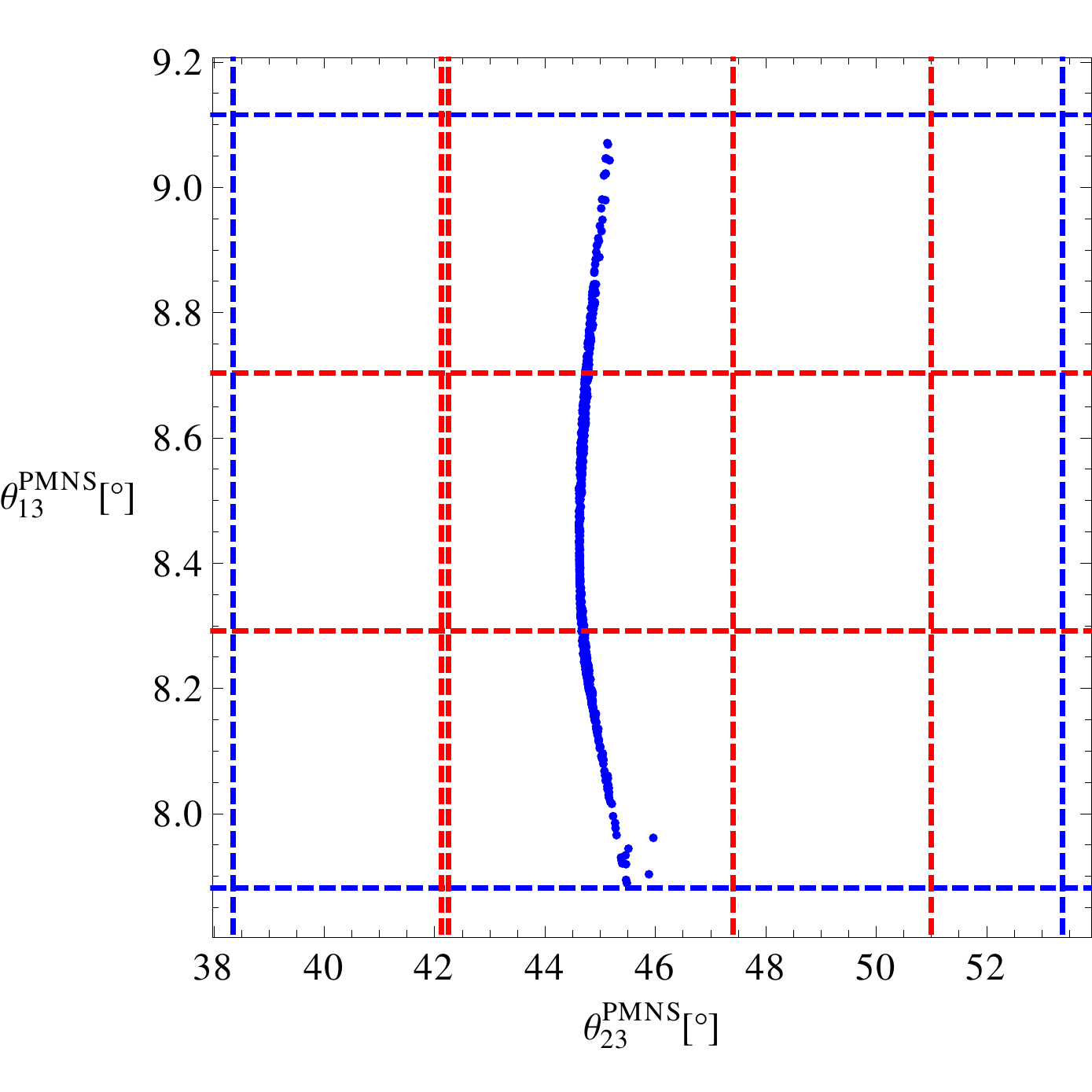} \hspace{0.7cm}
\includegraphics[scale=0.49]{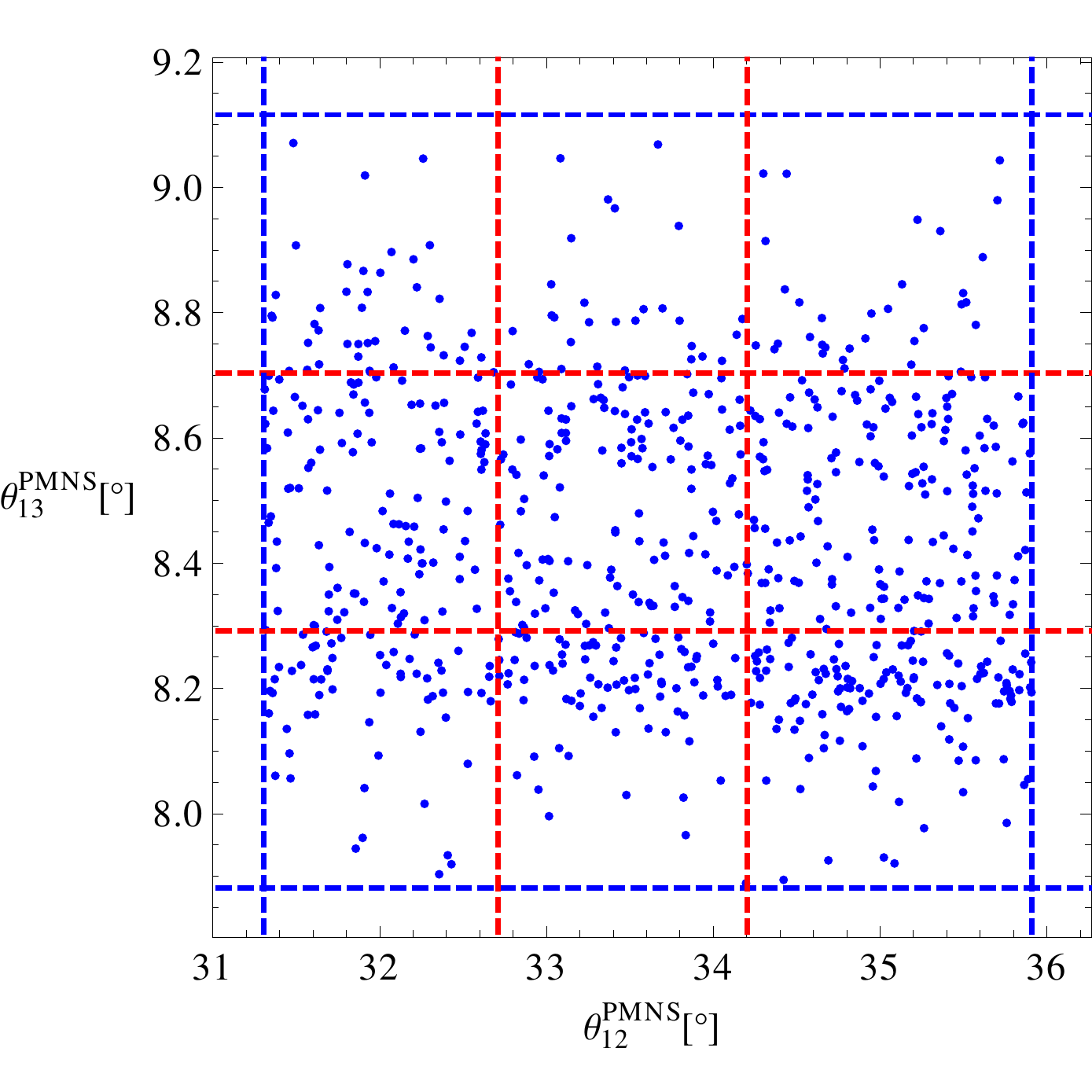}
\includegraphics[scale=0.49]{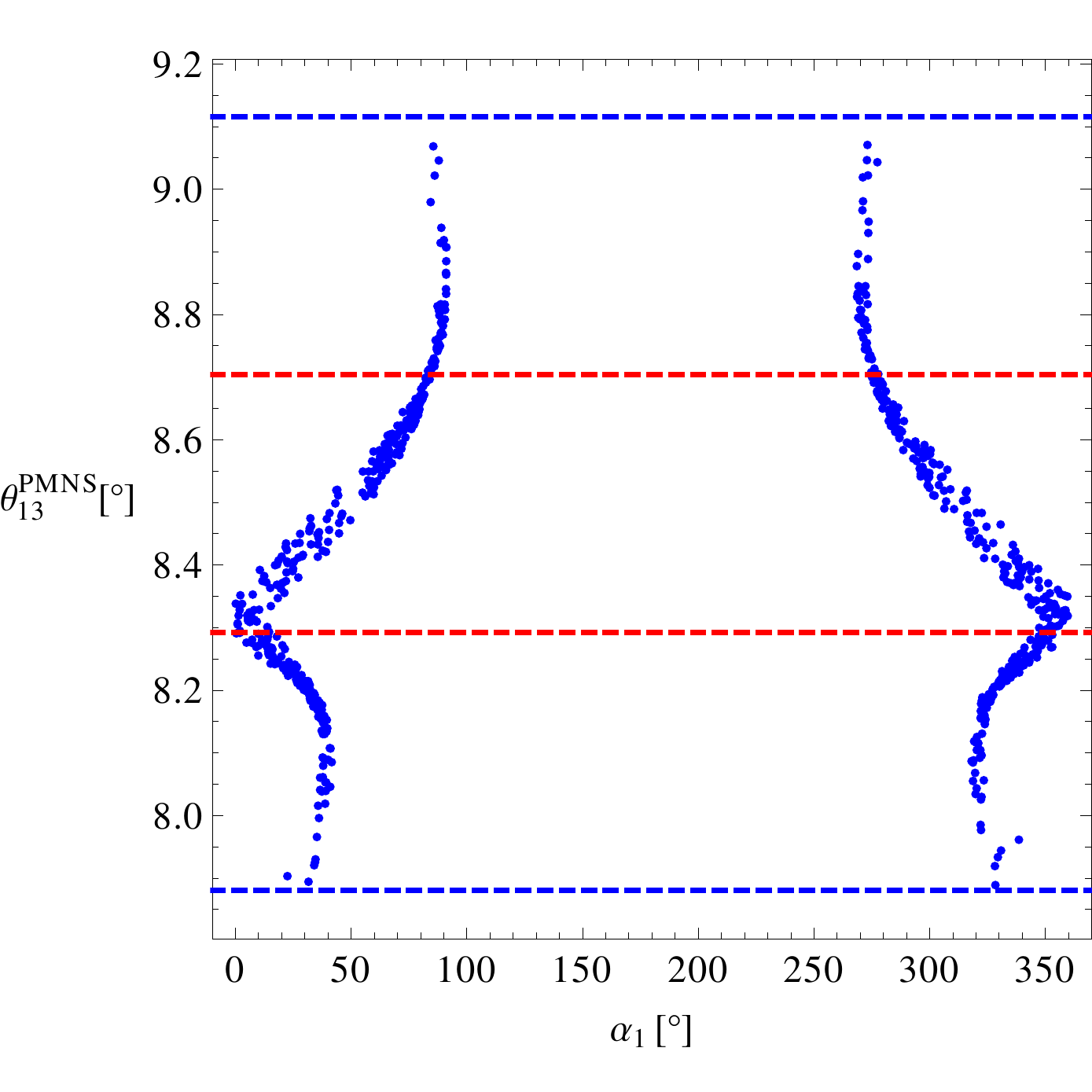} \hspace{0.7cm}
\includegraphics[scale=0.49]{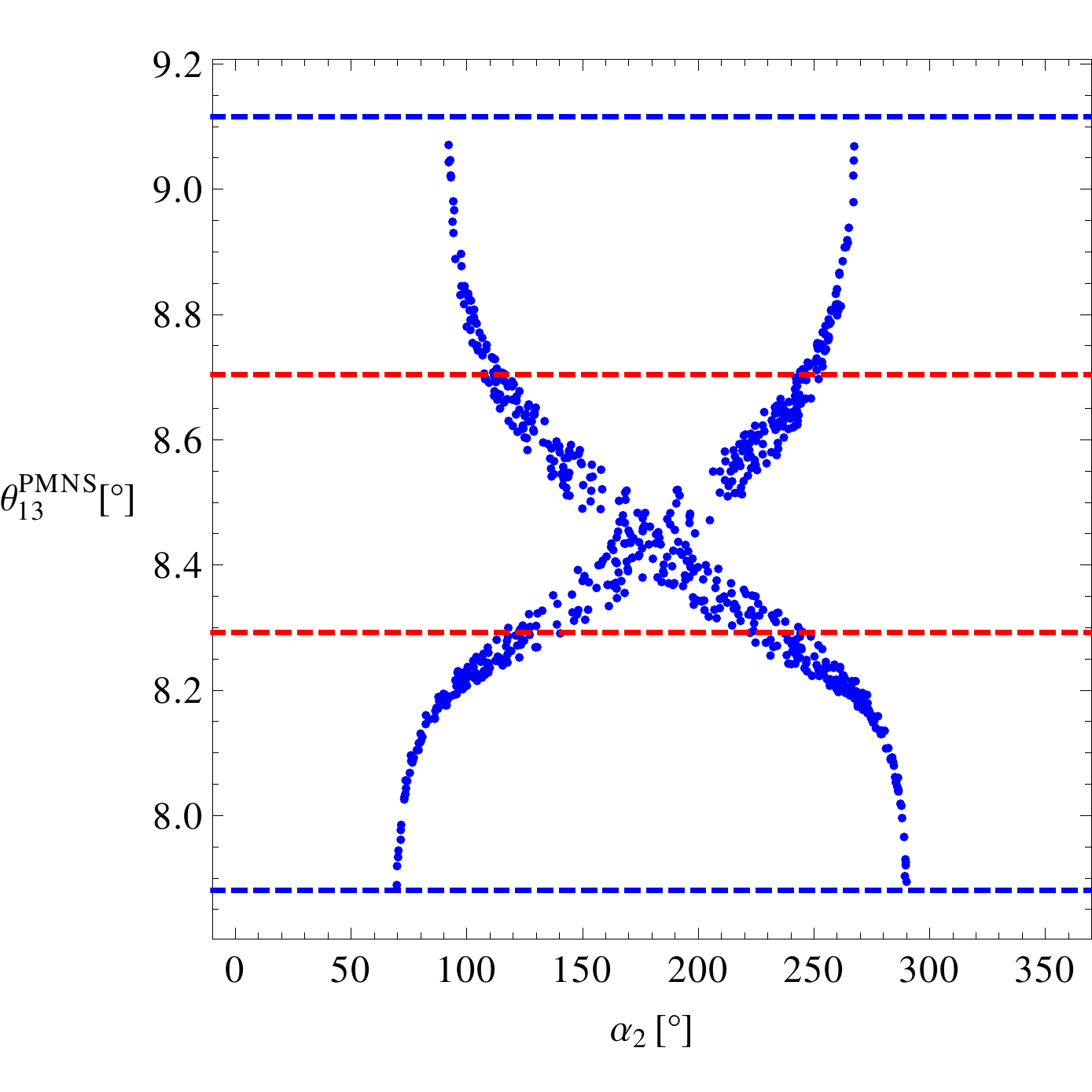}
\includegraphics[scale=0.49]{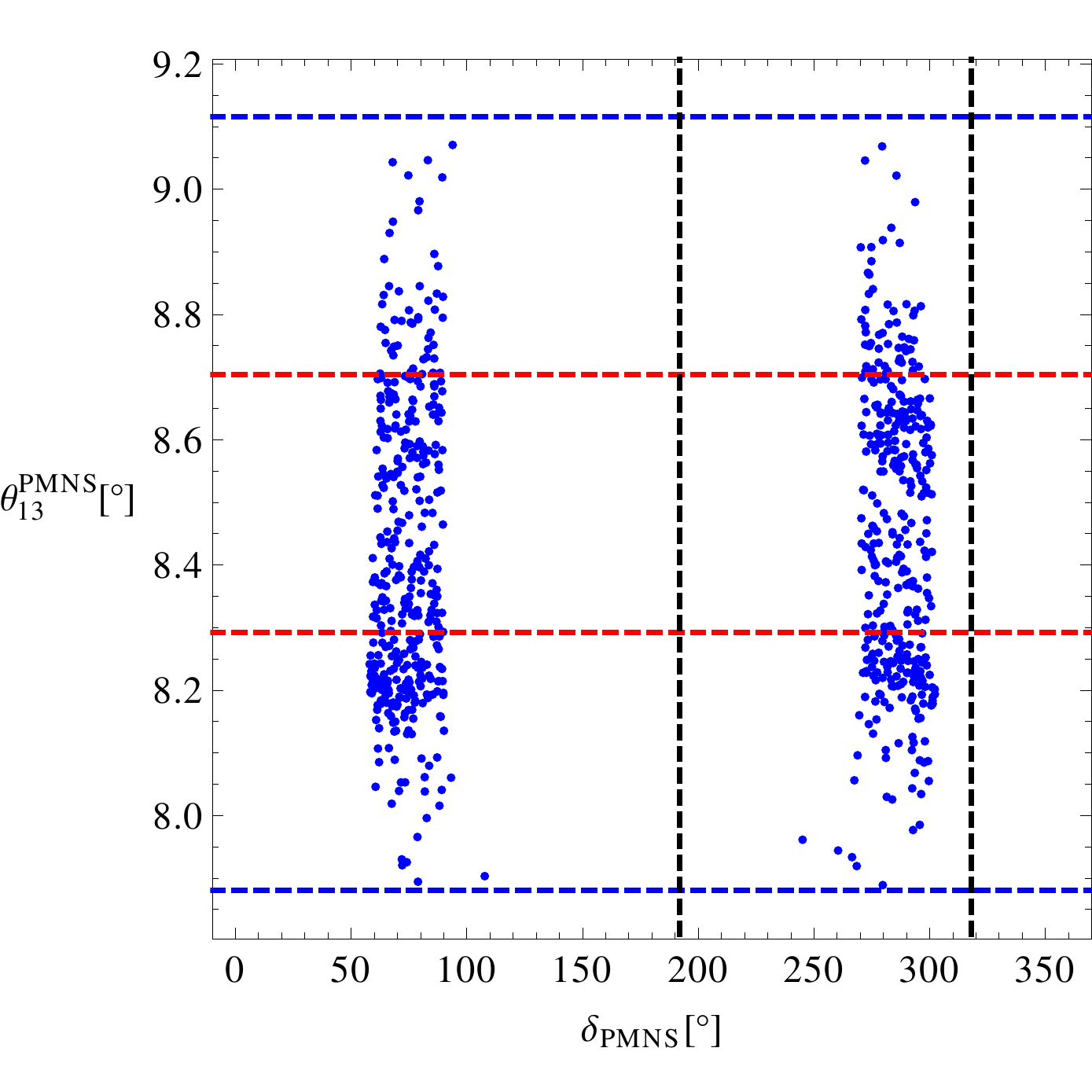} \hspace{0.7cm}
\includegraphics[scale=0.49]{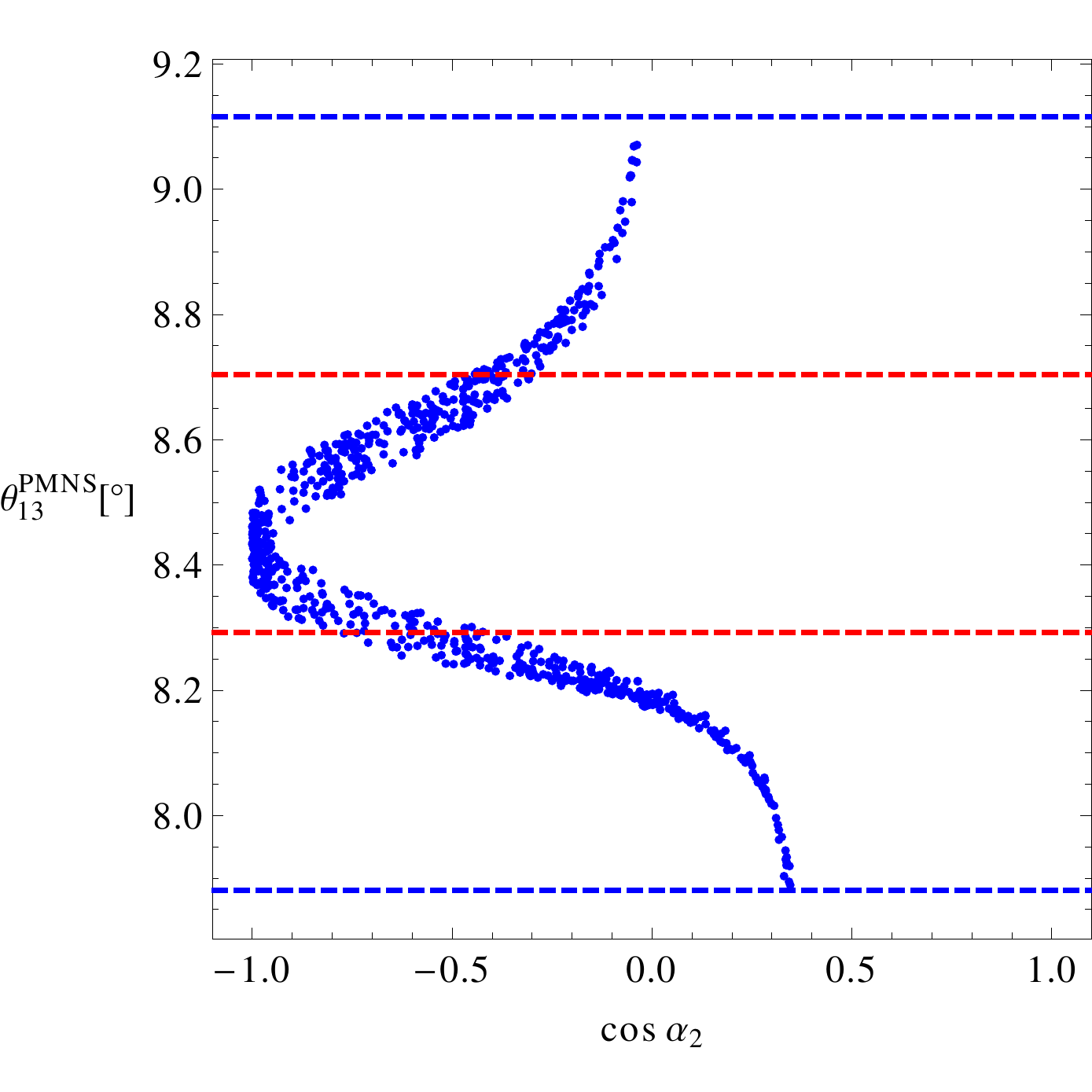}
\caption{
Results of our parameter scan for the normal hierarchy (blue points).
The allowed experimental 3$\sigma$ (1$\sigma$) regions are
limited by blue (red) dashed lines. The black dashed lines represent the 1$\sigma$
range for the not directly measured CP phase $\delta_{\text{PMNS}}$ from the global
fit \cite{nu-fit}.
}
\label{fig:mixing_parameters}
\end{figure}

Our numerical scan results for the leptonic mixing parameters are displayed in Fig.~\ref{fig:mixing_parameters},
where the allowed 3$\sigma$ (1$\sigma$) regions are limited by blue (red) dashed lines.
The black dashed lines represent the 1$\sigma$ range for the not directly measured CP phase
$\delta_{\text{PMNS}}$ from the global fit \cite{nu-fit}.
The blue points are the result from our parameter scan to which we have applied the experimental
data as constraints.

Note that $\theta_{23}^{\text{PMNS}}$ is not within the 1$\sigma$ region. And hence,
if it is confirmed that the atmospheric mixing is not close to maximal this concrete
model would be ruled out. Nevertheless, it is rather straightforward to introduce
a $\theta_{23}^e$ mixing which would allow to fit $\theta_{23}^{\text{PMNS}}$ but
would make the model much less predictive.

For the Majorana phases $\alpha_{1}$ and $\alpha_{2}$ we find values
between $0^{\circ}$--$90^{\circ}$ or $270^{\circ}$--$360^{\circ}$ for $\alpha_{1}$
and $70^{\circ}$--$290^{\circ}$ for $\alpha_{2}$. 
We find the Dirac phase $\delta_{\text{PMNS}}$ to be in the region from
$57^\circ$--$108^\circ$ or $244^\circ$--$303^\circ$.
The Jarlskog invariant which determines the CP violation in neutrino oscillations
is given by \cite{Jarlskog:1985ht}
\begin{equation}
J_{\text{CP}}=\text{Im}(U_{\mu 3}U_{e3}^{*}U_{e2}U_{\mu 2}^{*})=\frac{1}{8}\cos(\theta_{13}^{\text{PMNS}})\sin(2\theta_{12}^{\text{PMNS}})\sin(2\theta_{13}^{\text{PMNS}})\sin(2\theta_{23}^{\text{PMNS}})\sin{\delta_{\text{PMNS}}}.
\label{eq:jcpPDG}
\end{equation}
We obtain $J_{\text{CP}}\approx \pm (0.027 - 0.035)$.

We would like to mention here the work done in \cite{Petcov:2014laa} where among other things
a similar setup was studied and constraints for the phases were found. Nevertheless, the authors
neglected RGE running effects which they can do by assuming a small $\tan \beta$ or no supersymmetry
at all and furthermore they have no mass sum rule and therefore neutrino masses can be light in their
setup. Nevertheless, in the normal hierarchical setup where RGE effects do not have a large impact
we find similar results.

As we mentioned before the mass sum rule only implies a lower bound for the mass
scale for the normal hierarchy. But here we find as well an upper bound due to the
constraint that $\theta_{13}^{\text{PMNS}}$ should stay within the experimental
3$\sigma$ region. This can be clearly seen in the last plot in Fig.~\ref{fig:mixing_parameters}
where we have plotted  $\cos\left(\alpha_2\right)$ against $\theta_{13}^{\text{PMNS}}$. The
mass sum rule implies $\cos\left(\alpha_2\right)$ to be in the range from $-1$ to about $0.48$,
where larger values imply larger masses and larger RGE corrections to  $\th_{13}^{\text{PMNS}}$.

\begin{figure}
\includegraphics[scale=0.5]{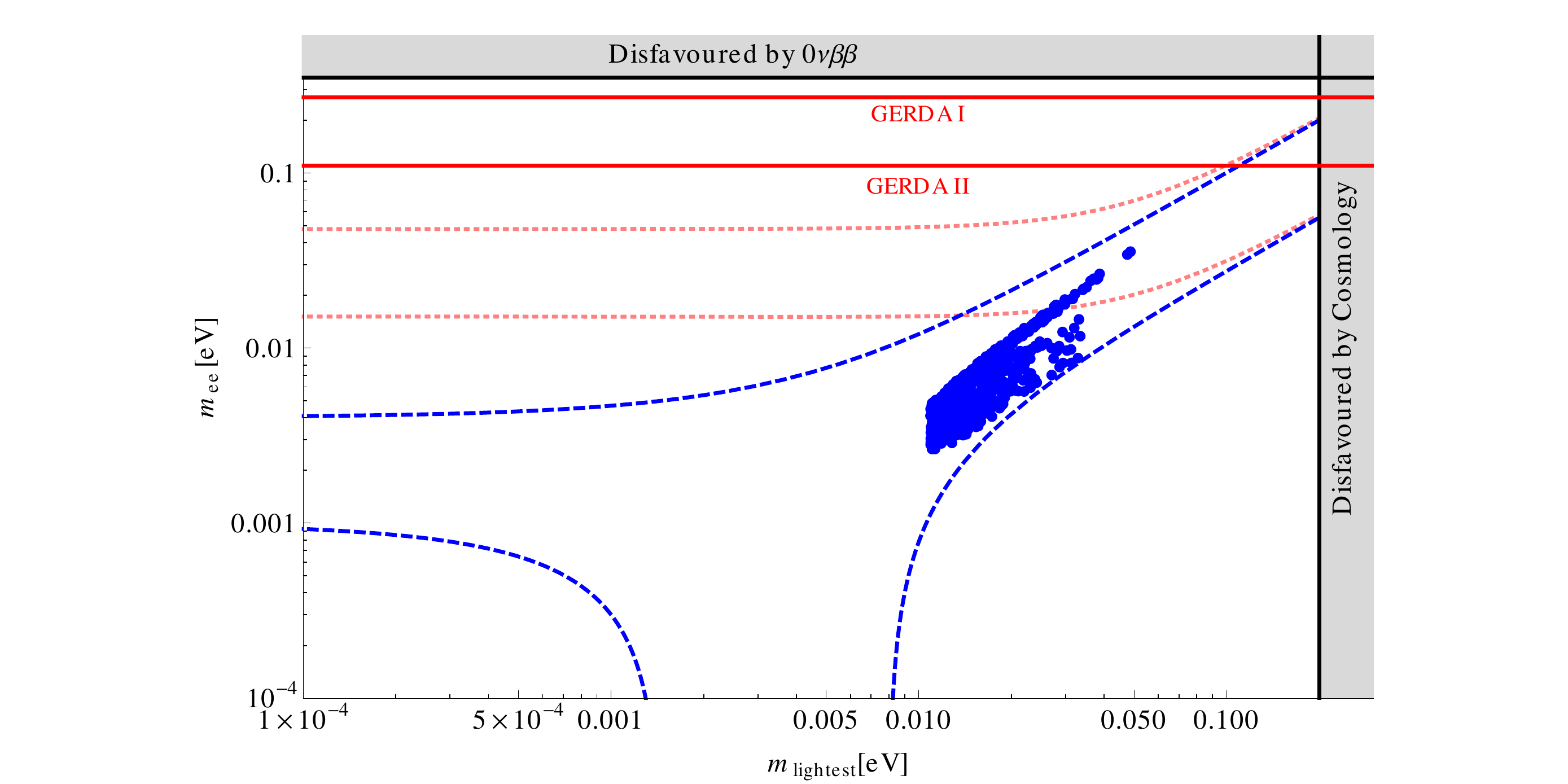}
\caption{
Prediction for the effective neutrino mass $m_{ee}$ accessible in neutrinoless double beta decay
experiments as a function of the lightest neutrino mass $m_{1}$. The blue dashed region represents the
allowed region for normal ordering whereas the pink dotted region indicates the inverted ordering
region which is not allowed in our setup. The grey region on the right side shows the bounds on the
lightest mass from cosmology \cite{Ade:2013zuv} and the grey region in the upper part displays the upper
bound on the effective mass from the EXO experiment \cite{Albert:2014awa}.
The red lines represent the sensitivity of GERDA phase I respectively GERDA phase II \cite{Smolnikov:2008fu}.
}
\label{fig:mee_plot}
\end{figure}

The effective neutrino mass accessible in neutrinoless double beta decay experiments like GERDA \cite{Smolnikov:2008fu}
or EXO \cite{Albert:2014awa} is given by 
\begin{equation}
\left|m_{ee}\right|=\left|m_{1} U_{e1}^{2}+m_{2} U_{e2}^{2}+m_{3} U_{e3}^{2}\right|=\left| m_{1}c_{12}^{2}c_{13}^{2}\text{e}^{-\text{i}\alpha_{1}}+m_{2}s_{12}^{2}c_{13}^{2}\text{e}^{-\text{i}\alpha_{2}}+m_{3}s_{13}^{2}\text{e}^{-\text{i}2\delta_{\text{PMNS}}}\right|.
\label{eq:mee}
\end{equation}
A graphically representation of our prediction for $m_{ee}$ as a function of $m_{1}$ is shown in
Fig.~\ref{fig:mee_plot}. We find values for $m_{ee}$ in the range from 0.02 eV to 0.04 eV 
corresponding to the lightest neutrino mass $m_{1}$  in the region from 0.01 eV to 0.05 eV.
This results are beyond the sensitivity of the GERDA experiment but might be tested by a future experiment.

With the value for the lightest neutrino mass $m_{1}$ between about 0.01~eV and 0.05~eV and the experimental mass squared
differences from Tab.~\ref{tab:exp_parameters} we obtain for the sum of the neutrino masses $\sum m_{\nu}= (0.074$--0.171)~eV.
This prediction is compatible with the cosmological bound for the sum of the neutrino masses \cite{Ade:2013zuv}
\begin{equation}
\sum m_{\nu} < 0.23 \text{ eV.}
\label{eq:kosmoschranke}
\end{equation} 

The quantity which will be measured in the experiment KATRIN \cite{Angrik:2005ep} is the kinematic neutrino mass
$m_{\beta}$ which is given as
\begin{equation}
m_{\beta}^{2}=m_{1}^{2}c_{12}^{2}c_{13}^{2}+m_{2}^{2}s_{12}^{2}c_{13}^{2}+m_{3}^{2}s_{13}^{2}.
\label{eq:katrin}
\end{equation}
Applying the range for $m_{1}$ as well as the measured mass squared differences we arrive at $m_{\beta} \approx (0.014$--$0.052)$~eV.
Regarding the sensitivity of the experiment which is $m_{\beta} > 0.2$~eV our model prediction is beyond the reach of KATRIN.

\section{Summary and Conclusions}
\label{sec:Summary}

In this paper we have presented the first SU(5)$\, \times \,$ A$_5$ SUSY Flavour Model to our knowledge.
It features to leading order the appealing prediction $\theta_{12}^{\text{PMNS}} =
\tan^{-1}\left(\frac{1}{\phi_{g}}\right)\approx 31.7^{\circ} $ where $\phi_g$ is the golden ratio $\phi_g = (1+\sqrt{5})/2$.  The
reactor mixing angle is predicted to be vanishing at leading order and the atmospheric mixing
angle to be maximal. Furthermore, the neutrino masses exhibit a sum rule, which turns out to
be very important for the phenomenology.

The prediction of a vanishing reactor mixing angle is excluded by several standard deviations
and hence the leading order predictions have to be corrected to make the model seem realistic.
In grand unified theories nevertheless, it is natural to expect that the charged lepton Yukawa
matrix is not diagonal because it is related to the down-type quark sector which is well motivated to be
non-diagonal in flavour space. This is furthermore suggested by the approximate relation $\theta_{13}^{\text{PMNS}}
\approx \theta_C/\sqrt{2}$, where $\theta_C$ is the Cabibbo angle. But in our setup we do not only
have relations between quark and lepton mixing angles, but also between down-type quark and charged
lepton Yukawa couplings which are non-standard, $y_\tau/y_b \approx -1.5$ and $y_\mu/y_s \approx 6$,
and for the double ratio $(y_\mu/y_s)(y_d/y_e) = 12$ which are all in perfect agreement with
experimental data. The Yukawa coupling ratios for the third and second generation put furthermore
two non-trivial constraints on the SUSY spectrum which might be tested at the LHC or one of its
successors.

To achieve the desired Yukawa coupling ratios and a non-diagonal charged lepton Yukawa matrix we have
presented a complete symmetry breaking sector for SU(5) and A$_5$.
The SU(5) breaking sector is peculiar because it is in principle compatible with the double missing
partner mechanism as discussed in \cite{Antusch:2014poa}, a mechanism to decouple the coloured
triplets and hence suppress proton decay sufficiently. In the A$_5$ symmetry breaking we have introduced
a few non-trivial representations which break A$_5$ in the desired groups such that we end up
with golden ratio mixing type A to leading order in the neutrino sector including also
a sum rule for the neutrino masses. We have also studied a messenger sector for the model which is
important for choosing between different Yukawa coupling relations in the effective higher-dimensional
operators and forbidding other unwanted effective operators which might be allowed by the symmetries
alone.

Apart from corrections from the charged lepton sector, RGE corrections can also play a major role.
In fact, RGE corrections rule out the inverted neutrino mass hierarchy. The neutrino mass sum
rule allows both mass hierarchies but in both cases only a certain mass range.
For inverted  hierarchy the neutrino masses turn out to be rather heavy and since $\tan \beta$ is as well
rather large the RGE corrections to $\theta_{12}^{\text{PMNS}}$ are so large that although at the high
scale we are at most a few degrees away from the observed value at low energies we are far outside
the allowed 3$\sigma$ region for $\theta_{12}^{\text{PMNS}}$.
Hence, only the normal hierarchy is
possible in our model setup and we find all three mixing angles to be in the 3$\sigma$ regions
and $J_{\text{CP}} \approx \pm 0.03$ with the lightest neutrino mass $m_1 \approx 0.01$--$0.05$~eV.
Due to the mass and angle sum rules we also find constraints on the phases, most phenomenologically
relevant for the near future, $\delta_{\text{PMNS}}$ to be in the region from $57^\circ$--$108^\circ$
or $244^\circ$--$303^\circ$.

Hence, our model can be tested from neutrino and collider experiments in several different ways in the
near future.

\section*{Note Added}

During the finalisation of this work an update of the nu-fit global
fitting collaboration appeared \cite{Gonzalez-Garcia:2014bfa}.
Nevertheless, the results which we used in our analysis changed
only very little compared to their updated fit and hence our conclusions
remain unchanged.

\section*{Acknowledgements}

We would like to thank A.~Meroni for sharing her code with us for generating
the $m_{ee}$ vs.~$m_{\text{lightest}}$ plot. JO acknowledges
partial support from the Vector Foundation and MS would like to thank
UGM Yogyakarta for kind hospitality during finishing of this work.

\appendix

\section{The messenger sector}
 \label{app:UV}

In this section we discuss the renormalisable superpotential of the model. 
As mentioned before the heavy messenger fields are integrated out to obtain
the higher dimensional operators of the effective superpotential.
The complete messenger field content can be found in
Tabs.~\ref{tab:messengerZn} and \ref{tab:messengerSinglet}.

\begin{table}
\centering
\begin{tabular}{l c l c c c c c c c c c c}
\toprule
& $\mathrm{SU(5)}$ & $\mathrm{A_5}$ & $\zz_{4}^R$ & $\zz_2$ & $\zz_2$ &
$\zz_3$ & $\zz_3$ & $\zz_3$ & $\zz_3$ & $\zz_3$ & $\zz_3$ & $\zz_4$ \\
\midrule
$\Sigma_1                           $ & $\mathbf{5}       $ & $\mathbf{3}$ & $1$ & $0$ & $0$ & $2$ & $1$ & $2$ & $1$ & $0$ & $1$ & $0$ \\
$\bar{\Sigma}_1                     $ & $\mathbf{\bar{5}} $ & $\mathbf{3}$ & $1$ & $0$ & $0$ & $1$ & $2$ & $1$ & $2$ & $0$ & $2$ & $0$ \\
$\Sigma_2                           $ & $\mathbf{5}       $ & $\mathbf{1}$ & $1$ & $0$ & $0$ & $2$ & $1$ & $1$ & $2$ & $0$ & $1$ & $3$ \\
$\bar{\Sigma}_2                     $ & $\mathbf{\bar{5}} $ & $\mathbf{1}$ & $1$ & $0$ & $0$ & $1$ & $2$ & $2$ & $1$ & $0$ & $2$ & $1$ \\
\midrule
$\Xi_1                              $ & $\mathbf{45}      $ & $\mathbf{3}$ & $1$ & $0$ & $0$ & $2$ & $1$ & $2$ & $1$ & $0$ & $1$ & $0$ \\
$\bar{\Xi}_1                        $ & $\mathbf{\bar{45}}$ & $\mathbf{3}$ & $1$ & $0$ & $0$ & $1$ & $2$ & $1$ & $2$ & $0$ & $2$ & $0$ \\
$\Xi_2                              $ & $\mathbf{45}      $ & $\mathbf{3}$ & $1$ & $0$ & $1$ & $2$ & $1$ & $1$ & $1$ & $2$ & $0$ & $1$ \\
$\bar{\Xi}_2                        $ & $\mathbf{\bar{45}}$ & $\mathbf{3}$ & $1$ & $0$ & $1$ & $1$ & $2$ & $2$ & $2$ & $1$ & $0$ & $3$ \\
$\Xi_3                              $ & $\mathbf{45}      $ & $\mathbf{1}$ & $1$ & $1$ & $0$ & $2$ & $2$ & $1$ & $2$ & $0$ & $0$ & $0$ \\
$\bar{\Xi}_3                        $ & $\mathbf{\bar{45}}$ & $\mathbf{1}$ & $1$ & $1$ & $0$ & $1$ & $1$ & $2$ & $1$ & $0$ & $0$ & $0$ \\
$\Xi_4                              $ & $\mathbf{45}      $ & $\mathbf{1}$ & $2$ & $0$ & $0$ & $0$ & $0$ & $1$ & $0$ & $0$ & $0$ & $0$ \\
$\bar{\Xi}_4                        $ & $\mathbf{\bar{45}}$ & $\mathbf{1}$ & $0$ & $0$ & $0$ & $0$ & $0$ & $2$ & $0$ & $0$ & $0$ & $0$ \\
\midrule
$\Gamma_1                           $ & $\mathbf{1}       $ & $\mathbf{3}$ & $0$ & $0$ & $0$ & $0$ & $1$ & $1$ & $1$ & $0$ & $0$ & $0$ \\
$\bar{\Gamma}_1                     $ & $\mathbf{1}       $ & $\mathbf{3}$ & $2$ & $0$ & $0$ & $0$ & $2$ & $2$ & $2$ & $0$ & $0$ & $0$ \\
$\Gamma_2                           $ & $\mathbf{1}       $ & $\mathbf{3}$ & $0$ & $0$ & $0$ & $0$ & $1$ & $2$ & $0$ & $0$ & $0$ & $1$ \\
$\bar{\Gamma}_2                     $ & $\mathbf{1}       $ & $\mathbf{3}$ & $2$ & $0$ & $0$ & $0$ & $2$ & $1$ & $0$ & $0$ & $0$ & $3$ \\
$\Gamma_3                           $ & $\mathbf{1}       $ & $\mathbf{1}$ & $0$ & $0$ & $0$ & $2$ & $2$ & $2$ & $0$ & $0$ & $0$ & $0$ \\
$\bar{\Gamma}_3                     $ & $\mathbf{1}       $ & $\mathbf{1}$ & $2$ & $0$ & $0$ & $0$ & $1$ & $1$ & $1$ & $0$ & $0$ & $0$ \\
\midrule
$\Omega_1                           $ & $\mathbf{10}      $ & $\mathbf{1}$ & $1$ & $0$ & $0$ & $0$ & $1$ & $0$ & $0$ & $0$ & $0$ & $3$ \\
$\bar{\Omega}_1                     $ & $\mathbf{\bar{10}}$ & $\mathbf{1}$ & $1$ & $0$ & $0$ & $0$ & $2$ & $0$ & $0$ & $0$ & $0$ & $1$ \\
$\Omega_2                           $ & $\mathbf{10}      $ & $\mathbf{1}$ & $1$ & $1$ & $0$ & $2$ & $0$ & $0$ & $0$ & $0$ & $0$ & $0$ \\
$\bar{\Omega}_2                     $ & $\mathbf{\bar{10}}$ & $\mathbf{1}$ & $1$ & $1$ & $0$ & $1$ & $0$ & $0$ & $0$ & $0$ & $0$ & $0$ \\
$\Omega_3                           $ & $\mathbf{10}      $ & $\mathbf{1}$ & $1$ & $1$ & $0$ & $2$ & $2$ & $0$ & $0$ & $0$ & $0$ & $3$ \\
$\bar{\Omega}_3                     $ & $\mathbf{\bar{10}}$ & $\mathbf{1}$ & $1$ & $1$ & $0$ & $1$ & $1$ & $0$ & $0$ & $0$ & $0$ & $1$ \\
$\Omega_4                           $ & $\mathbf{10}      $ & $\mathbf{1}$ & $1$ & $1$ & $0$ & $1$ & $1$ & $2$ & $2$ & $0$ & $0$ & $2$ \\
$\bar{\Omega}_4                     $ & $\mathbf{\bar{10}}$ & $\mathbf{1}$ & $1$ & $1$ & $0$ & $2$ & $2$ & $1$ & $1$ & $0$ & $0$ & $2$ \\
$\Omega_5                           $ & $\mathbf{10}      $ & $\mathbf{3}$ & $1$ & $0$ & $0$ & $1$ & $2$ & $0$ & $1$ & $0$ & $2$ & $0$ \\
$\bar{\Omega}_5                     $ & $\mathbf{\bar{10}}$ & $\mathbf{3}$ & $1$ & $0$ & $0$ & $2$ & $1$ & $0$ & $2$ & $0$ & $1$ & $0$ \\
$\Omega_6                           $ & $\mathbf{10}      $ & $\mathbf{1}$ & $1$ & $0$ & $0$ & $1$ & $1$ & $1$ & $1$ & $0$ & $2$ & $3$ \\
$\bar{\Omega}_6                     $ & $\mathbf{\bar{10}}$ & $\mathbf{1}$ & $1$ & $0$ & $0$ & $2$ & $2$ & $2$ & $2$ & $0$ & $1$ & $1$ \\
$\Omega_7                           $ & $\mathbf{10}      $ & $\mathbf{3}$ & $1$ & $0$ & $0$ & $1$ & $1$ & $2$ & $0$ & $0$ & $2$ & $0$ \\
$\bar{\Omega}_7                     $ & $\mathbf{\bar{10}}$ & $\mathbf{3}$ & $1$ & $0$ & $0$ & $2$ & $2$ & $1$ & $0$ & $0$ & $1$ & $0$ \\
\bottomrule
\end{tabular}
\caption{The $\mathbb{Z}_n$ charges, $\mathrm{SU(5)}$ and $\mathrm{A_5}$
  representations of the messenger fields for the Yukawa couplings.}
\label{tab:messengerZn}
\end{table}

\begin{table}
\centering
\begin{tabular}{l c l c c c c c c c c c c}
\toprule
 & $\mathrm{SU(5)}$ & $\mathrm{A_5}$ & $\mathbb{Z}_{4R}$ & $\mathbb{Z}_2$ & $\mathbb{Z}_2$ & $\mathbb{Z}_3$ & $\mathbb{Z}_3$ & $\mathbb{Z}_3$ & $\mathbb{Z}_3$ & $\mathbb{Z}_3$ & $\mathbb{Z}_3$ & $\mathbb{Z}_4$ \\
\midrule
$\Upsilon_{\epsilon2}         $ & $\mathbf{1}$ & $\mathbf{1}$ & $2$ & $0$ & $0$ & $0$ & $2$ & $0$ & $0$ & $0$ & $0$ & $2$ \\
$\bar{\Upsilon}_{\epsilon2}   $ & $\mathbf{1}$ & $\mathbf{1}$ & $0$ & $0$ & $0$ & $0$ & $1$ & $0$ & $0$ & $0$ & $0$ & $2$ \\
$\Upsilon_{\epsilon3}         $ & $\mathbf{1}$ & $\mathbf{1}$ & $2$ & $0$ & $0$ & $1$ & $2$ & $0$ & $0$ & $0$ & $0$ & $0$ \\
$\bar{\Upsilon}_{\epsilon3}   $ & $\mathbf{1}$ & $\mathbf{1}$ & $0$ & $0$ & $0$ & $2$ & $1$ & $0$ & $0$ & $0$ & $0$ & $0$ \\
$\Upsilon_{\epsilon4}         $ & $\mathbf{1}$ & $\mathbf{1}$ & $2$ & $0$ & $0$ & $2$ & $2$ & $2$ & $2$ & $0$ & $0$ & $2$ \\
$\bar{\Upsilon}_{\epsilon4}   $ & $\mathbf{1}$ & $\mathbf{1}$ & $0$ & $0$ & $0$ & $1$ & $1$ & $1$ & $1$ & $0$ & $0$ & $2$ \\
$\Upsilon_{\epsilon5}         $ & $\mathbf{1}$ & $\mathbf{1}$ & $2$ & $0$ & $0$ & $1$ & $0$ & $2$ & $2$ & $0$ & $0$ & $2$ \\
$\bar{\Upsilon}_{\epsilon5}   $ & $\mathbf{1}$ & $\mathbf{1}$ & $0$ & $0$ & $0$ & $2$ & $0$ & $1$ & $1$ & $0$ & $0$ & $2$ \\
$\Upsilon_{\theta1}           $ & $\mathbf{1}$ & $\mathbf{1}$ & $2$ & $0$ & $0$ & $0$ & $2$ & $2$ & $1$ & $1$ & $0$ & $0$ \\
$\bar{\Upsilon}_{\theta1}     $ & $\mathbf{1}$ & $\mathbf{1}$ & $0$ & $0$ & $0$ & $0$ & $1$ & $1$ & $2$ & $2$ & $0$ & $0$ \\
$\Upsilon_{\theta2}           $ & $\mathbf{1}$ & $\mathbf{1}$ & $2$ & $0$ & $0$ & $0$ & $2$ & $1$ & $2$ & $1$ & $0$ & $2$ \\
$\bar{\Upsilon}_{\theta2}     $ & $\mathbf{1}$ & $\mathbf{1}$ & $0$ & $0$ & $0$ & $0$ & $1$ & $2$ & $1$ & $2$ & $0$ & $2$ \\
$\Upsilon_{\theta3}           $ & $\mathbf{1}$ & $\mathbf{1}$ & $2$ & $0$ & $0$ & $0$ & $0$ & $1$ & $0$ & $1$ & $1$ & $2$ \\
$\bar{\Upsilon}_{\theta3}     $ & $\mathbf{1}$ & $\mathbf{1}$ & $0$ & $0$ & $0$ & $0$ & $0$ & $2$ & $0$ & $2$ & $2$ & $2$ \\
\midrule
$\Lambda_{\epsilon2}          $ & $\mathbf{1}$ & $\mathbf{1}$ & $2$ & $0$ & $0$ & $0$ & $1$ & $0$ & $0$ & $0$ & $0$ & $0$ \\
$\bar{\Lambda}_{\epsilon2}    $ & $\mathbf{1}$ & $\mathbf{1}$ & $0$ & $0$ & $0$ & $0$ & $2$ & $0$ & $0$ & $0$ & $0$ & $0$ \\
$\Lambda_{\epsilon4}          $ & $\mathbf{1}$ & $\mathbf{1}$ & $2$ & $0$ & $0$ & $1$ & $1$ & $1$ & $1$ & $0$ & $0$ & $0$ \\
$\bar{\Lambda}_{\epsilon4}    $ & $\mathbf{1}$ & $\mathbf{1}$ & $0$ & $0$ & $0$ & $2$ & $2$ & $2$ & $2$ & $0$ & $0$ & $0$ \\
$\Lambda_{\epsilon5}          $ & $\mathbf{1}$ & $\mathbf{1}$ & $2$ & $0$ & $0$ & $2$ & $0$ & $1$ & $1$ & $0$ & $0$ & $0$ \\
$\bar{\Lambda}_{\epsilon5}    $ & $\mathbf{1}$ & $\mathbf{1}$ & $0$ & $0$ & $0$ & $1$ & $0$ & $2$ & $2$ & $0$ & $0$ & $0$ \\
$\Lambda_{\theta2}            $ & $\mathbf{1}$ & $\mathbf{1}$ & $2$ & $0$ & $0$ & $0$ & $1$ & $2$ & $1$ & $2$ & $0$ & $0$ \\
$\bar{\Lambda}_{\theta2}      $ & $\mathbf{1}$ & $\mathbf{1}$ & $0$ & $0$ & $0$ & $0$ & $2$ & $1$ & $2$ & $1$ & $0$ & $0$ \\
$\Lambda_{\theta3}            $ & $\mathbf{1}$ & $\mathbf{1}$ & $2$ & $0$ & $0$ & $0$ & $0$ & $2$ & $0$ & $2$ & $2$ & $0$ \\
$\bar{\Lambda}_{\theta3}      $ & $\mathbf{1}$ & $\mathbf{1}$ & $0$ & $0$ & $0$ & $0$ & $0$ & $1$ & $0$ & $1$ & $1$ & $0$ \\
\midrule
$\Delta_{\epsilon2}           $ & $\mathbf{1}$ & $\mathbf{1}$ & $2$ & $0$ & $0$ & $0$ & $2$ & $0$ & $0$ & $0$ & $0$ & $0$ \\
$\bar{\Delta}_{\epsilon2}     $ & $\mathbf{1}$ & $\mathbf{1}$ & $0$ & $0$ & $0$ & $0$ & $1$ & $0$ & $0$ & $0$ & $0$ & $0$ \\
$\Delta_{\epsilon3}           $ & $\mathbf{1}$ & $\mathbf{1}$ & $2$ & $0$ & $0$ & $2$ & $1$ & $0$ & $0$ & $0$ & $0$ & $0$ \\
$\bar{\Delta}_{\epsilon3}     $ & $\mathbf{1}$ & $\mathbf{1}$ & $0$ & $0$ & $0$ & $1$ & $2$ & $0$ & $0$ & $0$ & $0$ & $0$ \\
$\Delta_{\epsilon4}           $ & $\mathbf{1}$ & $\mathbf{1}$ & $2$ & $0$ & $0$ & $2$ & $2$ & $2$ & $2$ & $0$ & $0$ & $0$ \\
$\bar{\Delta}_{\epsilon4}     $ & $\mathbf{1}$ & $\mathbf{1}$ & $0$ & $0$ & $0$ & $1$ & $1$ & $1$ & $1$ & $0$ & $0$ & $0$ \\
$\Delta_{\epsilon5}           $ & $\mathbf{1}$ & $\mathbf{1}$ & $2$ & $0$ & $0$ & $1$ & $0$ & $2$ & $2$ & $0$ & $0$ & $0$ \\
$\bar{\Delta}_{\epsilon5}     $ & $\mathbf{1}$ & $\mathbf{1}$ & $0$ & $0$ & $0$ & $2$ & $0$ & $1$ & $1$ & $0$ & $0$ & $0$ \\
$\Delta_{\theta1}             $ & $\mathbf{1}$ & $\mathbf{1}$ & $2$ & $0$ & $0$ & $0$ & $1$ & $1$ & $2$ & $2$ & $0$ & $0$ \\
$\bar{\Delta}_{\theta1}       $ & $\mathbf{1}$ & $\mathbf{1}$ & $0$ & $0$ & $0$ & $0$ & $2$ & $2$ & $1$ & $1$ & $0$ & $0$ \\
$\Delta_{\theta2}             $ & $\mathbf{1}$ & $\mathbf{1}$ & $2$ & $0$ & $0$ & $0$ & $2$ & $1$ & $2$ & $1$ & $0$ & $0$ \\
$\bar{\Delta}_{\theta2}       $ & $\mathbf{1}$ & $\mathbf{1}$ & $0$ & $0$ & $0$ & $0$ & $1$ & $2$ & $1$ & $2$ & $0$ & $0$ \\
$\Delta_{\theta3}             $ & $\mathbf{1}$ & $\mathbf{1}$ & $2$ & $0$ & $0$ & $0$ & $0$ & $1$ & $0$ & $1$ & $1$ & $0$ \\
$\bar{\Delta}_{\theta3}       $ & $\mathbf{1}$ & $\mathbf{1}$ & $0$ & $0$ & $0$ & $0$ & $0$ & $2$ & $0$ & $2$ & $2$ & $0$ \\
\bottomrule
\end{tabular}
\caption{The $\mathbb{Z}_n$ charges, $\mathrm{SU(5)}$ and $\mathrm{A_5}$
  representations of the messenger fields for the flavon sector.}
\label{tab:messengerSinglet}
\end{table}

We will first discuss the renormalisable superpotentials for the up- and
down-type quark sectors including additional operators not seen in our
supergraphs but allowed by symmetry. We will then do the same for the flavon sector. At last we will
discuss higher dimensional operators.

We begin with the mass terms for the messenger fields
\begin{align}
\begin{split}
\mathcal{W}_\Lambda^{\text{ren}} &= M_{\Sigma_i}\Sigma_i \bar{\Sigma}_i +
M_{\Omega_i}\Omega_i \bar{\Omega_i} + M_{\Xi_i} \Xi_i \bar{\Xi}_i +
M_{\Gamma_i} \Gamma_i \bar{\Gamma}_i + M_{\Upsilon_{f6}} \Upsilon_{f6}
\bar{\Upsilon}_{f6} \\
 &+ M_{\Delta_{f6}} \Delta_{f6} \bar{\Delta}_{f6} + M_{\Upsilon_{f12}} \Upsilon_{f12}
\bar{\Upsilon}_{f12} + M_{\Delta_{f12}} \Delta_{f12} \bar{\Delta}_{f12} +
M_{\Lambda_{f12}} \Lambda_{f12} \bar{\Lambda}_{f12} \;, 
\end{split}
\end{align}
where a summation over $i$ is implied. The indices $f6$ and $f12$ denote the
singlet flavons which occur as 6th and 12th power respectively in their aligning superpotentials. It is $f6 \in \{\theta_1,\eps_3 \}$ and $f12 \in \{\theta_2,
\theta_3,\eps_2, \eps_4, \eps_5 \}$ where a summation over these flavons is implied. Each messenger field has a mass higher than
the GUT scale. The individual messenger masses are related
to the messenger mass scale $\Lambda$ by order one coefficients which are often
not explicitly stated to simplify the notation. 

\begin{figure}
\centering
\includegraphics[scale = 0.6]{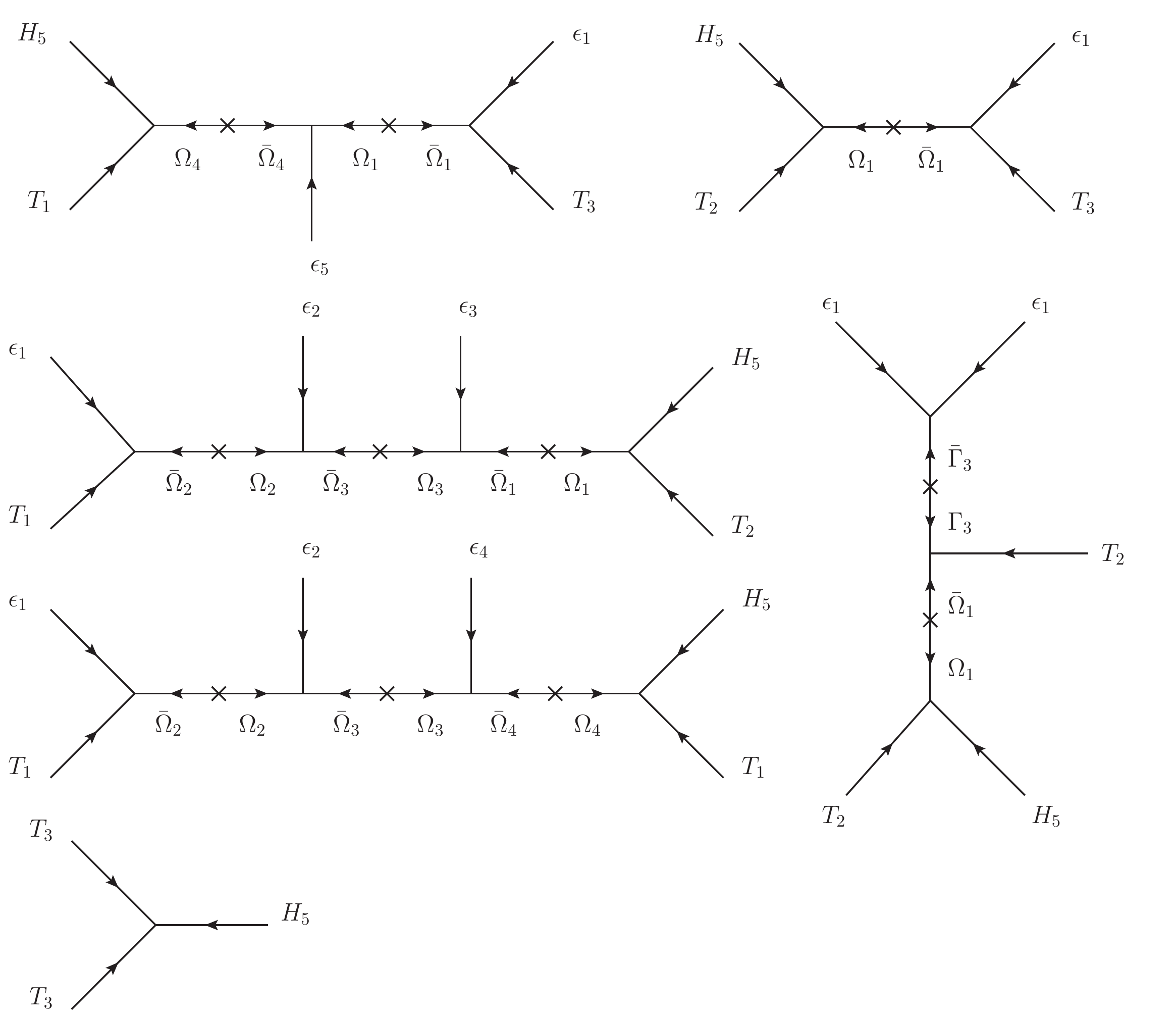}
\caption{The supergraphs before integrating out the heavy messenger field for
  the up-type quark sector. 
    }
\label{fig:upSector}
\end{figure}

The renormalisable superpotential for the up-quark sector is
\begin{align}
\begin{split}
  \mathcal{W}_u^{\text{ren}} &= T_3 T_3 H_5 + T_3 \epsilon_1 \bar{\Omega}_1 +
\Omega_1 T_2 H_5 + \epsilon_1^2 \bar{\Gamma}_3 + \Gamma_3 T_2 \bar{\Omega}_1 + \epsilon_1 T_1 \bar{\Omega}_2 \\
& + \Omega_2 \epsilon_2 \bar{\Omega}_3 
+ \Omega_3 \bar{\Omega}_1 \epsilon_3 + \Omega_3 \bar{\Omega}_4 \epsilon_4
+ \Omega_4 T_1 H_5 + \bar{\Omega}_4 \Omega_1 \epsilon_5 \;,
\end{split}
\label{eq:renWu}
\end{align} 
where the coupling constants have been ommitted to increase clarity.
The supergraphs for this sector can be found in Fig.~\ref{fig:upSector}.
In order to get the effective operators in section \ref{sec:model}, the
messenger fields have to be integrated out.

\begin{figure}
\centering
\includegraphics[scale = 0.56]{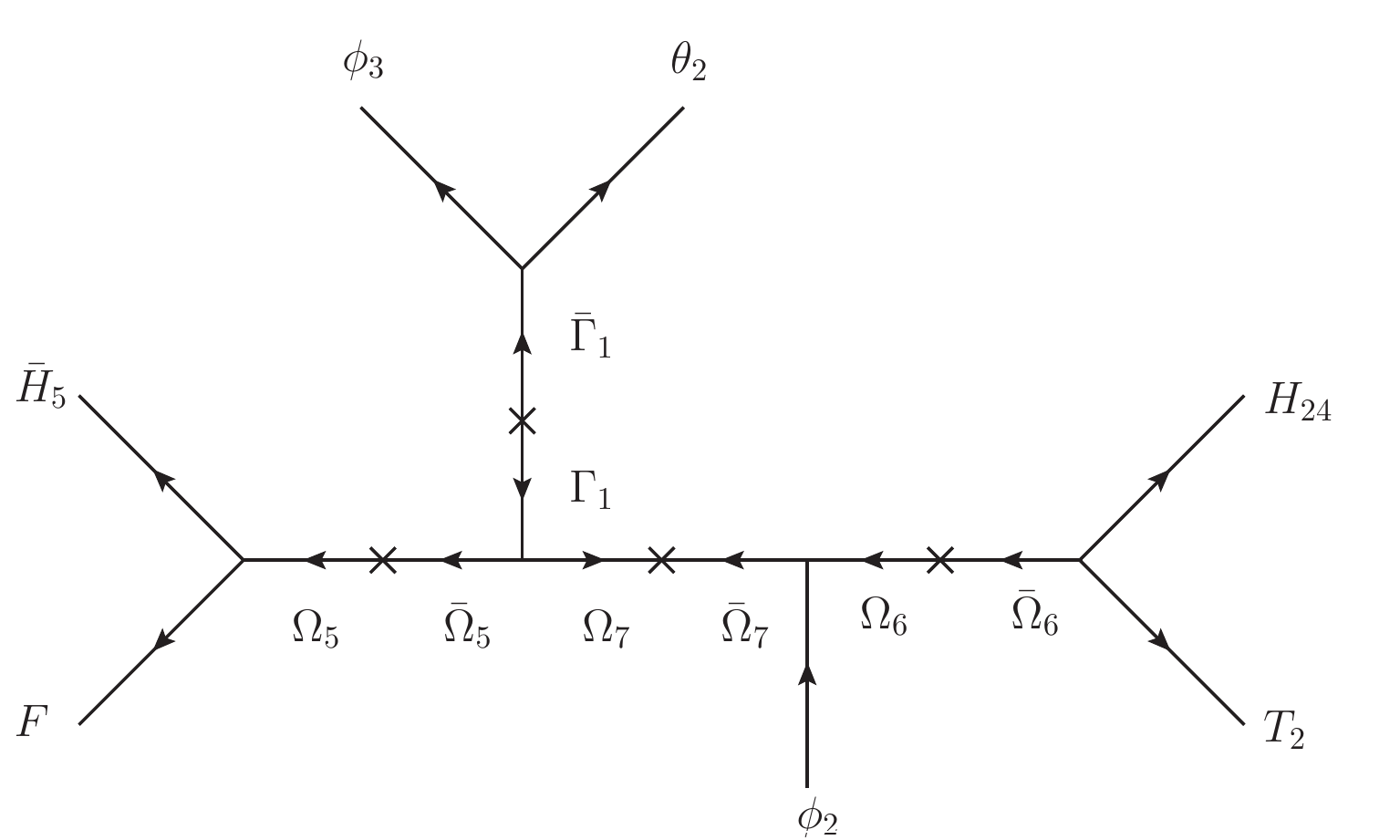}
\includegraphics[scale = 0.56]{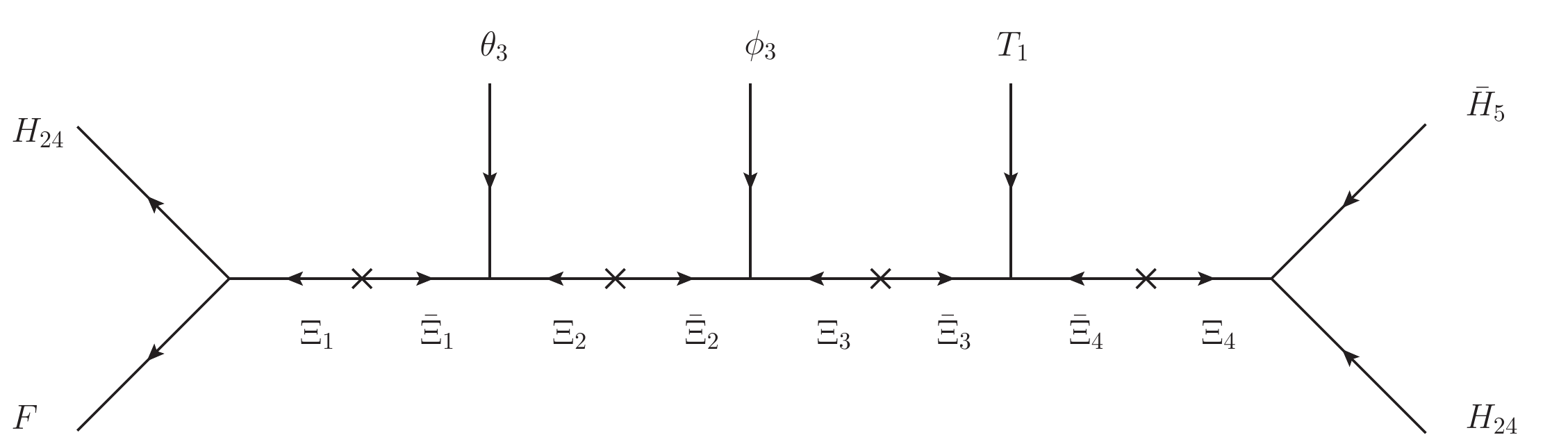}
\includegraphics[scale = 0.56]{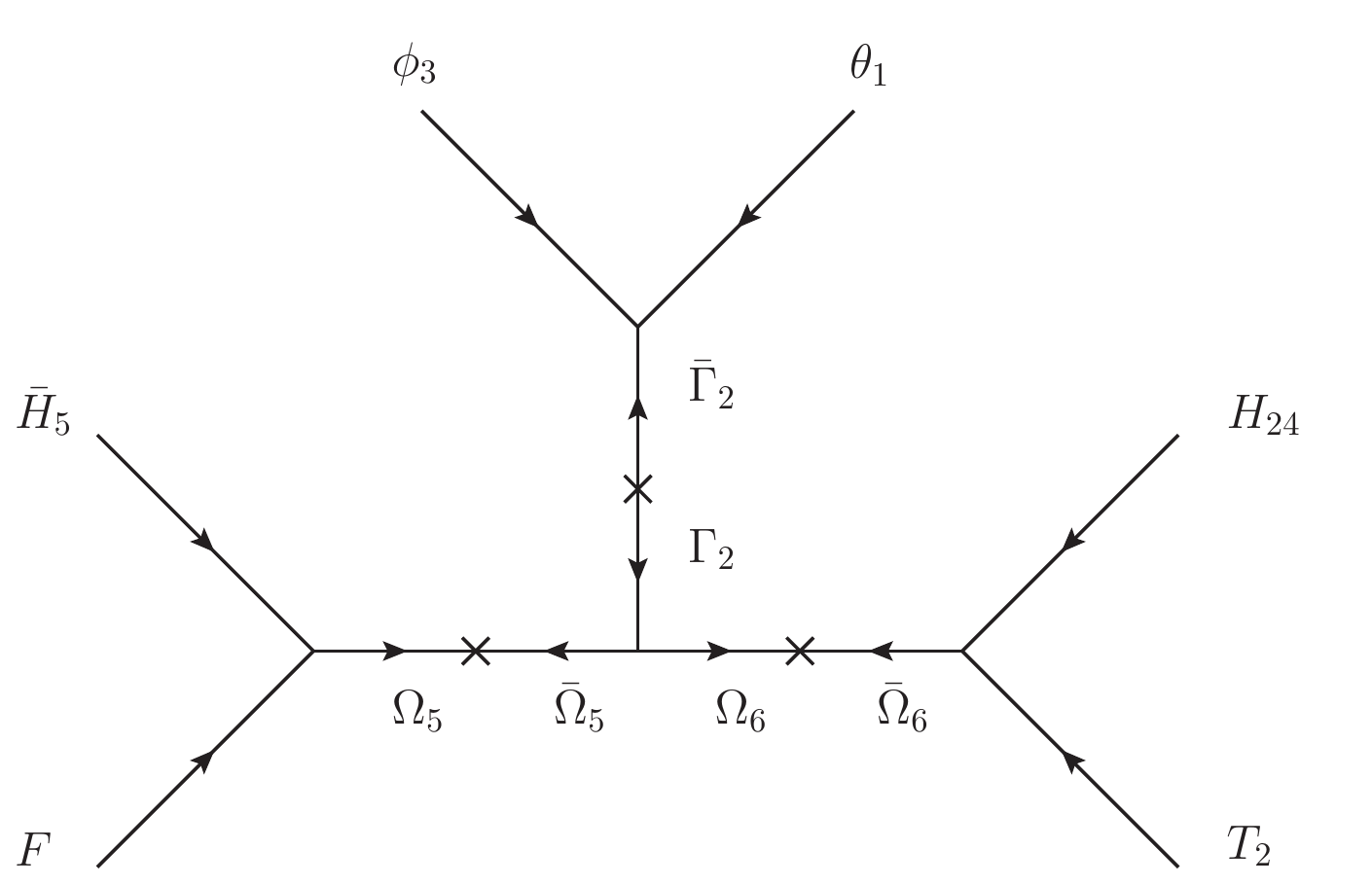}
\includegraphics[scale = 0.56]{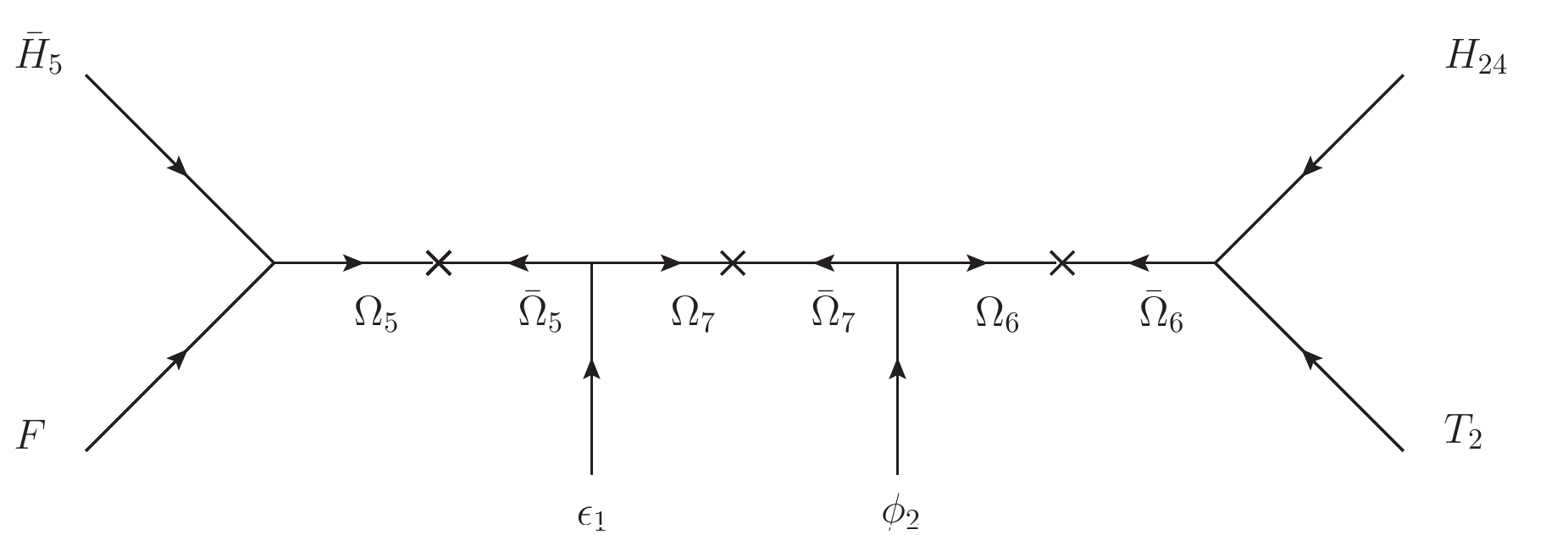}
\includegraphics[scale = 0.56]{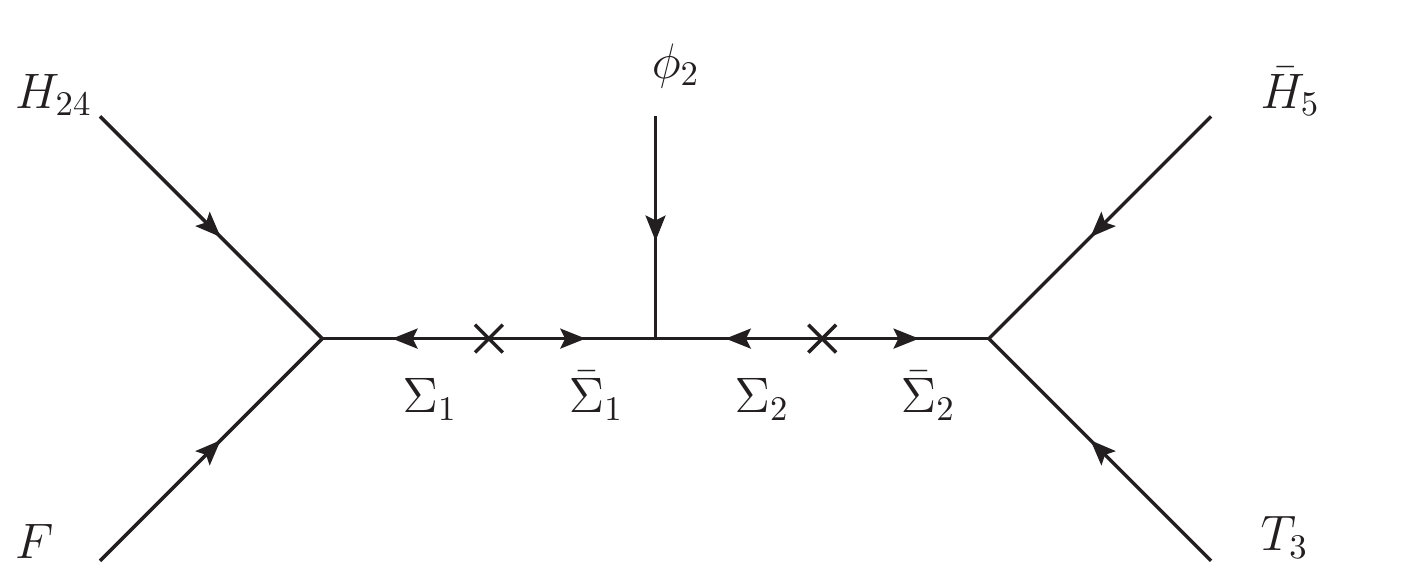}
\caption{The supergraphs for the down-type quark and charged lepton
  sector.
  }
\label{fig:downSector}
\end{figure}

The renormalisable superpotential for the charged lepton and down-type quark sector
is 
\begin{align}
\begin{split}
\mathcal{W}_{d,l}^{\text{ren}} &= H_{24}  F  \Sigma_1 + \bar{\Sigma}_1 \phi_2 \Sigma_2 +
\bar{\Sigma}_2 T_3 \bar{H}_5 
+ F H_{24} \Xi_1 + \bar{\Xi}_1 \theta_3 \Xi_2 \\ 
&+  \bar{\Xi}_2 \phi_3 \Xi_3 + \bar{\Xi}_3 \bar{\Xi}_4 T_1 + \Xi_4 \bar{H}_5 H_{24}
\\
 &+ F \bar{H}_5  \Omega_5 + \bar{\Omega}_5 \Omega_6 \Gamma_2 + \bar{\Omega}_6
H_{24} T_2 + \bar{\Gamma}_2 \phi_3 \theta_1 \\
&+ \bar{\Omega}_5 \Omega_7 \Gamma_1 + \bar{\Gamma}_1 \theta_2 \phi_3 +
\bar{\Omega}_7 \Omega_6 \phi_2 
+ \bar{\Omega}_5 \Omega_7 \epsilon_1 \;,  
\end{split}
\label{eq:renWdl}
\end{align}
where again coupling constants have been omitted.
The charges under the shaping symmetries are listed in
Tab.~\ref{tab:messengerZn} 
for the messenger fields and Tab.~\ref{tab:matterZn} for the matter and Higgs
fields of the model.
The supergraphs for this sector can be found in Fig.~\ref{fig:downSector}.
There are a few additional couplings which are not forbidden by shaping
symmetries. These are
\begin{align}
\mathcal{W}_{\text{additional}} &= \bar{\Gamma}_3^3 + \bar{\Gamma}_3 \Gamma_1^2 +
\Gamma_3 \Omega_5 \bar{\Omega}_7 + \Gamma_1 \bar{\Gamma}_2 \f_2 +
\bar{\Gamma}_2 \epsilon_1 \phi_2 \;. 
\end{align}
It is important to note that the vertices above which contain $\Gamma_3$ or
$\epsilon_1$ and any messenger field of the down-sector are the only
allowed couplings that mix messenger fields of the up- and down-sector. We
will discuss the implications of these terms on potential higher dimensional
operators later. 

The operator $\bar{\Gamma}_2 \phi_2 \epsilon_1$ generates a second leading
order diagram
for the 3-2 element of $Y_d$ (and the 2-3 element of $Y_e$ respectively). Since it
generates the same effective operator as the supergraph shown in Fig.~\ref{fig:downSector} with
the same CG coefficient in the charged lepton sector, we have omitted the
diagram. The same reasoning applies to the term $\Gamma_1 \bar{\Gamma}_2
\phi_2$ which generates a second leading order diagram for the 1-2 element of
$Y_d$ (and the 2-1 element of $Y_e$ respectively).

There are more couplings between the messenger fields of the singlet
flavons. These will be further discussed below, since there are no couplings
mixing the singlet messenger fields with messenger fields from any other sector.

\begin{figure}
\centering
\includegraphics[scale = 0.56]{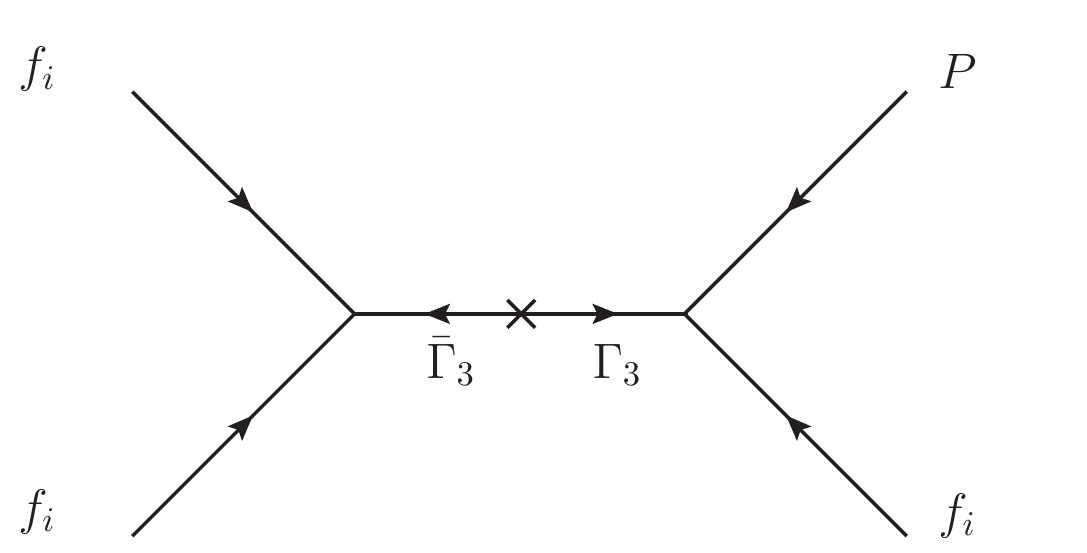}
\includegraphics[scale = 0.56]{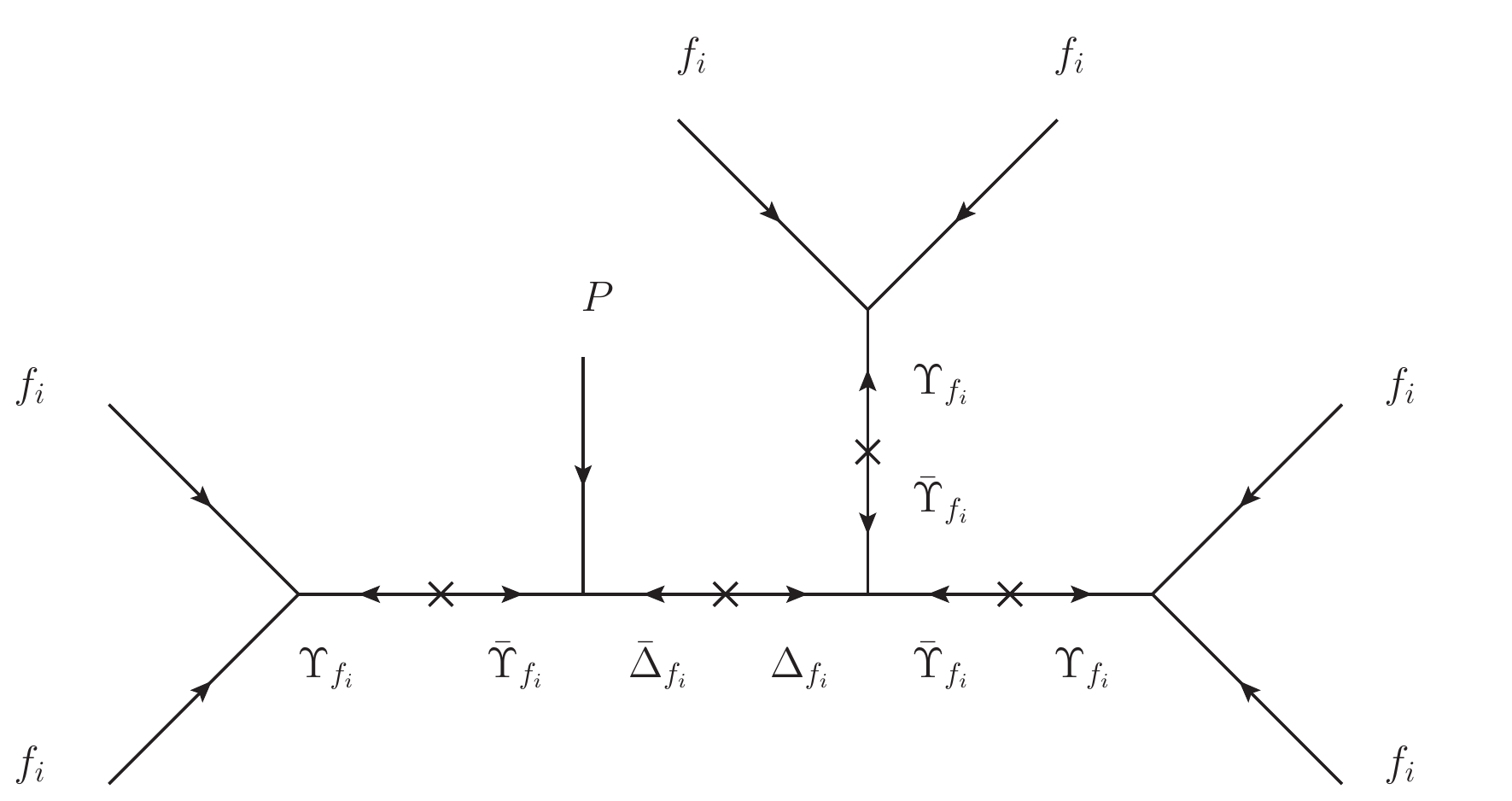}
\includegraphics[width=\textwidth]{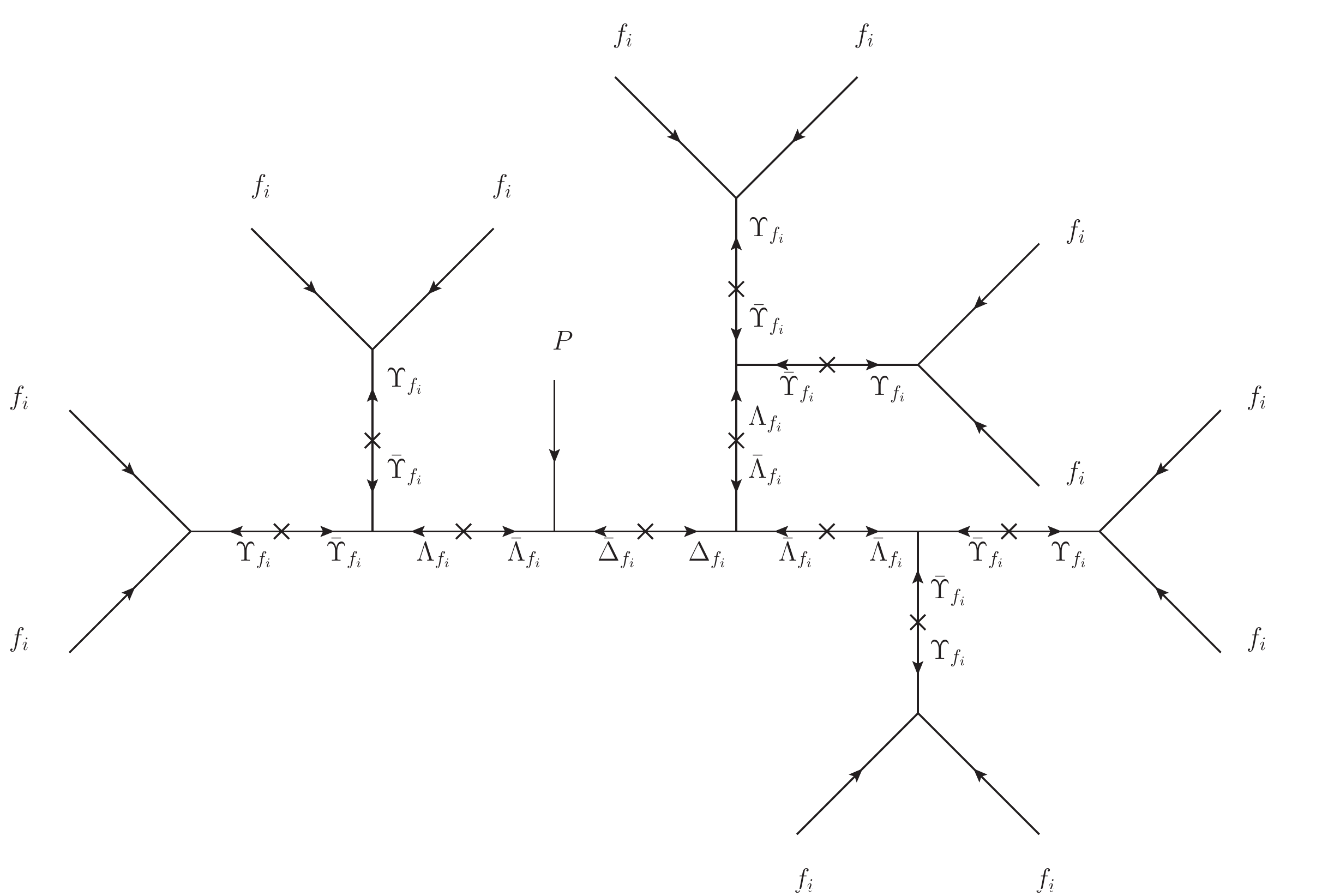}
\caption{The supergraphs for the sector of the singlet flavons.}
\label{fig:singlets}  
\end{figure}

In the flavon sector only the singlet alignment requires the introduction of
new messenger fields, since the superpotential for the flavon fields in the three-,
four- and five-dimensional representations of $\mathrm{A}_5$ is already renormalisable.
The messenger fields for the flavon sector and their charges under the
various symmetries of the model can be found in Tab.~\ref{tab:messengerSinglet}. 
For $\epsilon_1$ the renormalisable superpotential reads
\begin{align}
\mathcal{W}_{s3}^{\text{ren}} &= \epsilon_1^2 \bar{\Gamma}_3 + P \epsilon_1 \Gamma_3 \; .  
\end{align}
The renormalisable superpotentials of $\epsilon_3$ and $\theta_1$ are of the
form
\begin{align}
\mathcal{W}_{s6}^{\text{ren}} &= f_i^2 \Upsilon_{fi} + \bar{\Upsilon}_{fi}^2 \Delta_{fi} + P
\bar{\Delta}_{fi} \bar{\Upsilon}_{fi} \; ,
\end{align}
where $f_i$ denotes one of the above mentioned flavons. After integrating out
the heavy messenger fields we get an effective operator which contains the
respective flavon to the power of six.
The remaining singlets have superpotentials of the form
\begin{align}
\mathcal{W}_{s12}^{\text{ren}} &= f_i^2 \Upsilon_{fi} + \bar{\Upsilon}_{fi}^2 \Lambda_{fi} +
 \bar{\Lambda}_{fi}^2 \Delta_{fi} + P \bar{\Delta}_{fi} \bar{\Lambda}_{fi} \; ,
\end{align}
where $f_i$ again denotes one of the mentioned flavons. This superpotential
results in an effective operator containing the singlet flavon to the power of
12. The corresponding supergraphs can be found in Fig.~\ref{fig:singlets}.

As already discussed in the matter sector there are additional couplings
among the messenger fields of the flavon sector not forbidden by symmetry. However,
note that these messenger fields do not couple to any other
sector with the exception of one term which will be discussed in detail
later. These terms will not be displayed here, since they do not lead to new
leading order effective operators. We have checked this already on the
effective level without resorting to messenger selection rules.

The only non-trivial operator left to discuss is $D_{\phi}^ {(2)} \Upsilon_{\theta_2}
\Delta_{\theta_1}$ which generates an effective operator $D_{\phi}^ {(2)} P^2 \theta_1^2 \theta_2^{10}$
which nevertheless, due to the vanishing of $\langle P \rangle$, does not have any effect whatsoever.

We turn now to the additional effective operators for the Yukawa matrices. It is useful to
recall their structure here to leading order.
In the down-type quark and charged lepton sector we have
\begin{align}
Y_d^{\text{LO}} &= \begin{pmatrix}
0 & \Lambda^{-4} & 0\\
\Lambda^{-4} & \Lambda^{-3} & 0\\
0 & \Lambda^{-3} & \Lambda^{-2}\\
\end{pmatrix}\;,
\label{eq:orderYd}
\end{align}
 and in the up-type quark sector
\begin{align}
Y_u^{\text{LO}} &= \begin{pmatrix}
\Lambda^{-3} & \Lambda^{-3} & \Lambda^{-2}\\
\Lambda^{-3} &\Lambda^{-2} & \Lambda^{-1}\\
\Lambda^{-2} & \Lambda^{-1} & 1\\
\end{pmatrix}\;.
\label{eq:orderYu}
\end{align}
For both sectors we have checked for possible additional effective operators using only the
symmetries of the model, i.e.~without considering messenger fields. 
In the up-type quark sector the largest corrections come from operators with a mass
dimension at least two higher than the leading order operator. We therefore have
$Y_u = Y_u^{\text{LO}} + Y_u^{\text{HO}}$, where 
\begin{align}
Y_u^{\text{HO}} &\lesssim \begin{pmatrix}
\Lambda^{-6} & \Lambda^{-6} & \Lambda^{-5}\\
\Lambda^{-6} & \Lambda^{-5} & \Lambda^{-4} \\  
\Lambda^{-5} & \Lambda^{-4} & \Lambda^{-3}\\
\end{pmatrix} \; .
\end{align}
Hence, we can neglect them.

In the down-type quark sector we have as well calculated higher order effective operators
based on symmetry arguments only, where operators containing $\phi_2^2$ or
$\phi_3^2$ where ignored,
because $\langle \phi_2 \rangle^2 = \langle \phi_3 \rangle^2 = 0$. We find five additional effective
operators
\begin{align}
\begin{split}
\mathcal{W}^{\text{HO}} &=  \frac{1}{\Lambda^4} H_{24} \bar{H}_5 F T_1 \phi_2 \eps_3 \eps_2 +
  \frac{1}{\Lambda^4} H_{24} \bar{H}_5 F T_1 \phi_2 \eps_5 \eps_1 +
  \frac{1}{\Lambda^4} H_{24} \bar{H}_5 F T_1 \phi_3 \eps_5 \theta_1 \\
& +  \frac{1}{\Lambda^5} H_{24} \bar{H}_5 F T_1 \phi_2 \phi_3 \theta_2 \eps_5 +
  \frac{1}{\Lambda^5} H_{24} \bar{H}_5 F T_3 \phi_3 \theta_1 \eps_1 \eps_1\;.
\end{split}
\label{eq:Who}
\end{align} 
Upon close inspection of the terms in eq.~\eqref{eq:Who} it becomes clear, that
those terms are forbidden due to messenger arguments. As stated above there are
no couplings other than to $\Gamma_3$ and $\eps_1$ that mix up-type quark and
down-type quark  messenger fields. Since $\eps_5$ and $\eps_3$ do not immediately
couple to $\Gamma_3$ it is impossible to generate the terms containing only
those flavons, since further external legs from the up-sector would arise.
The last term in eq.~\eqref{eq:Who} cannot be
realised since $T_3$ couples only to $\bar{\Sigma}_2$, a messenger field in the
$\mathbf{\bar{5}}$ representation of SU(5). There are no couplings mixing this
messenger field with any of the fields in other representations of SU(5), making
the effective operator containing $T_3$ and $\eps_1$ impossible. 

In conclusion we found for the down-type Yukawa matrix $Y_d = Y_d^{\text{LO}} +
Y_d^{\text{HO}}$
\begin{align}
Y_d^{\text{HO}} &\lesssim \begin{pmatrix}
\Lambda^{-6} & \Lambda^{-6} & \Lambda^{-6}\\
\Lambda^{-6} & \Lambda^{-6} & \Lambda^{-6} \\  
\Lambda^{-6} & \Lambda^{-6} & \Lambda^{-5}\\
\end{pmatrix}\;,
\end{align}
which can again be safely neglected.

\section{A$_{\mathbf{5}}$ Clebsch-Gordan Coefficients}

For convenience we give here the Clebsch-Gordan coefficients of the
group A$_5$, taken from \cite{Cooper:2012bd}. 
We use the notation $a_i$ ($b_i$) for elements of the first (second)
representation. The subscript $a$ ($s$) denotes antisymmetric (symmetric)
representations.

\begin{center}
\begin{tabular}{|c|c|}
\hline
$\bo{3} \ot \bo{3} = \bo{1_s} \op \bo{3_a} \op \bo{5_s}$ & $\bo{3^\prime} \ot
\bo{3^\prime} = \bo{1_s} \op \bo{3_a^\prime} \op \bo{5_s}$ \\
\hline
\hline
   &\\
$\bo{1_s} \sim a_1 b_1 + a_2 b_3 + a_3 b_2$ & $\bo{1_s} \sim a_1 b_1 + a_2 b_3
+ a_3 b_2$ \\  &\\

\begin{math}
\bo{3_a} \sim \begin{pmatrix}
a_2 b_3 - a_3 b_2\\
a_1 b_2 - a_2 b_1\\
a_3 b_1 - a_1 b_3\\
\end{pmatrix}
\end{math} &
 \begin{math}
\bo{3_a}^\prime \sim \begin{pmatrix}
a_2 b_3 - a_3 b_2\\
a_1 b_2 - a_2 b_1\\
a_3 b_1 - a_1 b_3\\
\end{pmatrix}
\end{math}\\  & \\

\begin{math}
\bo{5_s} \sim \begin{pmatrix}
2 a_1 b_1 - a_2 b_3-a_3 b_2\\
-\sqrt{3} a_1 b_2 - \sqrt{3} a_2 b_1\\
\sqrt{6} a_2 b_2\\
\sqrt{6} a_3 b_3 \\
-\sqrt{3} a_1 b_3 - \sqrt{3} a_3 b_1\\
\end{pmatrix}
\end{math} &

\begin{math}
\bo{5_s} \sim \begin{pmatrix}
2 a_1 b_1 - a_2 b_3-a_3 b_2\\
\sqrt{6} a_3 b_3 \\
-\sqrt{3} a_1 b_2 - \sqrt{3} a_2 b_1\\
-\sqrt{3} a_1 b_3 - \sqrt{3} a_3 b_1\\
\sqrt{6} a_2 b_2\\
\end{pmatrix}
\end{math}\\ & \\
\hline
\end{tabular}
\end{center}

\begin{center}
\begin{tabular}{|c|}
\hline
 $\bo{3} \ot
\bo{3^\prime} =  \bo{4} \op \bo{5}$ \\
\hline
\hline
  \\
\begin{math}
\bo{4} \sim \begin{pmatrix}
a_3 b_2 + \sqrt{2} a_2 b_1\\
- a_3 b_3 -\sqrt{2} a_1 b_2\\
-a_2 b_2 -\sqrt{2} a_1 b_3\\
a_2 b_3 + \sqrt{2} a_3 b_1\\
\end{pmatrix}
\end{math} 
 
\begin{math}
\bo{5} \sim \begin{pmatrix}
\sqrt{3} a_1 b_1\\
-\sqrt{2} a_3 b_2 + a_2 b_1 \\
-\sqrt{2} a_3 b_3 +a_1 b_2 \\
-\sqrt{2} a_2 b_2 +a_1 b_3\\
 a_3 b_1-\sqrt{2} a_2 b_3\\
\end{pmatrix}
\end{math}\\  \\
\hline
\end{tabular}
\end{center}

\begin{center}
\begin{tabular}{|c|c|}
\hline
$\bo{3} \ot \bo{4} =  \bo{3^\prime} \op \bo{4} \op \bo{5}$ & $\bo{3^\prime} \ot
\bo{4} = \bo{3} \op \bo{4} \op \bo{5}$ \\
\hline
\hline
 &\\
\begin{math}
\bo{3^\prime} \sim \begin{pmatrix}
-\sqrt{2} (a_2 b_4 + a_3 b_1)\\
\sqrt{2} a_1 b_2 - a_2 b_1 + a_3 b_3\\
\sqrt{2} a_1 b_3 + a_2 b_2 - a_3 b_4\\ 
\end{pmatrix}
\end{math} &
\begin{math}
\bo{3} \sim \begin{pmatrix}
-\sqrt{2} (a_2 b_3 + a_3 b_2)\\
\sqrt{2} a_1 b_1 + a_2 b_4 - a_3 b_3\\
\sqrt{2} a_1 b_4  - a_2 b_2 + a_3 b_1\\ 
\end{pmatrix}
\end{math}\\  &\\

\begin{math}
\bo{4} \sim \begin{pmatrix}
a_1 b_1 - \sqrt{2} a_3 b_2\\
- a_1 b_2 -\sqrt{2} a_2 b_1\\
a_1 b_3 +\sqrt{2} a_3 b_4\\
-a_1 b_4 + \sqrt{2} a_2 b_3\\
\end{pmatrix}
\end{math} &
 
\begin{math}
\bo{4} \sim \begin{pmatrix}
a_1 b_1 + \sqrt{2} a_3 b_3\\
 a_1 b_2 -\sqrt{2} a_3 b_4\\
-a_1 b_3 +\sqrt{2} a_2 b_1\\
-a_1 b_4 - \sqrt{2} a_2 b_2\\
\end{pmatrix}
\end{math}\\  &\\
 
\begin{math}
\bo{5} \sim \begin{pmatrix}
\sqrt{6}( a_2 b_4 - a_3 b_1)\\
\sqrt{2}2 a_1 b_1 + 2a_3 b_2 \\
-\sqrt{2} a_1 b_2 +a_2 b_1 +3 a_3 b_3\\
\sqrt{2} a_1 b_3 - 3a_2 b_2 - a_3 b_4\\
-2\sqrt{2} a_1 b_4-2a_2 b_3\\
\end{pmatrix}
\end{math}&

\begin{math}
\bo{5} \sim \begin{pmatrix}
\sqrt{6} (a_2 b_3 - a_3 b_2)\\
\sqrt{2} a_1 b_1 - 3 a_2 b_4 - a_3 b_3\\
2\sqrt{2} a_1 b_2 + 2 a_3 b_4\\
-2 \sqrt{2} a_1 b_3 - 2 a_2 b_1\\
-\sqrt{2} a_1b_4 + a_2 b_2 + 3 a_3 b_1\\
\end{pmatrix}
\end{math} \\& \\
\hline
\end{tabular}
\end{center}

\begin{center}
\begin{tabular}{|c|c|}
\hline
$\bo{3} \ot \bo{5} = \bo{3} \op \bo{3^\prime} \op \bo{4} \op \bo{5}$ &
$\bo{3^\prime} \ot \bo{5} = \bo{3^\prime} \op \bo{3} \op \bo{4} \op \bo{5}$ \\
\hline
\hline
 & \\
\begin{math}
\bo{3} \sim \begin{pmatrix}
-2 a_1 b_1 +\sqrt{3}a_2 b_5 + \sqrt{3}a_3 b_2\\
\sqrt{3}a_1 b_2 + a_2 b_1 -\sqrt{6}a_3 b_3\\
\sqrt{3}a_1 b_5 -\sqrt{6}a_2 b_4 + a_3 b_1\\
\end{pmatrix}
\end{math} &

\begin{math}
\bo{3} \sim \begin{pmatrix}
 a_2 b_4 +\sqrt{3}a_1 b_1 + a_3 b_3\\
-\sqrt{2}a_2 b_5 + a_1 b_2 -\sqrt{2}a_3 b_4\\
-\sqrt{2}a_3 b_2 -\sqrt{2}a_2 b_3 + a_1 b_5\\
\end{pmatrix}
\end{math} \\&  \\
\begin{math}
\bo{3^\prime} \sim \begin{pmatrix}
\sqrt{3}a_1 b_1 + a_2 b_5 + a_3 b_2 \\
a_1 b_3 -\sqrt{2} a_2 b_2 -\sqrt{2} a_3 b_4\\
a_1 b_4 -\sqrt{2}(a_2 b_3 + a_3 b_5)\\ 
\end{pmatrix}
\end{math} & 

\begin{math}
\bo{3^\prime} \sim \begin{pmatrix}
-2 a_1 b_1 +\sqrt{3}a_2 b_4 + \sqrt{3}a_3 b_3\\
\sqrt{3}a_1 b_3 + a_2 b_1 -\sqrt{6}a_3 b_5\\
\sqrt{3}a_1 b_4 -\sqrt{6}a_2 b_2 + a_3 b_1\\
\end{pmatrix}
\end{math}\\&  \\

\begin{math}
\bo{4} \sim \begin{pmatrix}
a_3 b_3- \sqrt{6}a_2 b_1 +2 \sqrt{2} a_1 b_2\\
- 3 a_3 b_4 -\sqrt{2} a_1 b_3 + 2 a_2 b_2\\
3 a_2 b_3 +\sqrt{2} a_1 b_4 - 2a_3 b_5\\
-a_2 b_4 - 2\sqrt{2} a_1 b_5 + \sqrt{6} a_3 b_1\\
\end{pmatrix}
\end{math} &
\begin{math}
\bo{4} \sim \begin{pmatrix}
 3 a_2 b_5 +\sqrt{2} a_1 b_2 - 2 a_3 b_4\\
a_3 b_5- \sqrt{6}a_2 b_1 +2 \sqrt{2} a_1 b_3\\
 -a_2 b_2 - 2\sqrt{2} a_1 b_4 + \sqrt{6} a_3 b_1\\
-3 a_3 b_2 -\sqrt{2} a_1 b_5 + 2a_2 b_3\\
\end{pmatrix}
\end{math}\\  &\\
 
\begin{math}
\bo{5} \sim \begin{pmatrix}
\sqrt{3}(a_2 b_5 - a_3 b_2)\\
-a_1 b_2 -\sqrt{3} a_2 b_1 - \sqrt{2} a_3 b_3\\
-2 a_1 b_3 -\sqrt{2}a_2 b_2\\
2 a_1 b_4 + \sqrt{2}a_3 b_5 \\
a_1 b_5 + \sqrt{2}a_2 b_4 + \sqrt{3}a_3 b_1\\
\end{pmatrix}
\end{math}&

\begin{math}
\bo{5} \sim \begin{pmatrix}
\sqrt{3}(a_2 b_4 - a_3 b_3)\\
2 a_1 b_2 +\sqrt{2}a_3 b_4\\
-a_1 b_3 -\sqrt{3} a_2 b_1 - \sqrt{2} a_3 b_5\\
a_1 b_4 + \sqrt{2}a_2 b_2 + \sqrt{3}a_3 b_1\\
-2 a_1 b_5 - \sqrt{2}a_2 b_3 \\
\end{pmatrix}
\end{math} \\ & \\
\hline
\end{tabular}
\end{center}

\begin{center}
\begin{tabular}{|c|c|}
\hline
$\bo{4} \ot \bo{4} = \bo{1_s} \op \bo{3_a^\prime} \op \bo{3_a} \op \bo{4_s} \op \bo{5_s}$ & $\bo{4} \ot
\bo{5} = \bo{3^\prime} \op \bo{3} \op \bo{4} \op \bo{5_1} \op \bo{5_2}$ \\
\hline
\hline
 & \\
$\bo{1_s} \sim a_1 b_4 + a_2 b_3 + a_3 b_2 + a_4 b_1$ &
\begin{math}
\bo{3} \sim \begin{pmatrix}
2 \sqrt{2}(a_1 b_5 -a_4 b_2) + \sqrt{2}(a_3 b_3-a_2 b_4)\\
-\sqrt{6}a_1b_1 + 2 a_2 b_5 + 3 a_3 b_4 - a_4 b_3 \\
a_1 b_4 - 3a_2 b_3 -2 a_3 b_2 + \sqrt{6} a_4 b_1\\
\end{pmatrix}
\end{math}\\ & \\

\begin{math}
\bo{3_a} \sim \begin{pmatrix}
 -a_1b_4 + a_2 b_3 -a_3 b_2 +a_4 b_1\\
\sqrt{2}(a_2 b_4 - a_4 b_2)\\
\sqrt{2}(a_1 b_3 -a_3 b_1)\\
\end{pmatrix}
\end{math} &
\begin{math}
\bo{3^\prime} \sim \begin{pmatrix}
 \sqrt{2}(a_1 b_5 -a_4 b_2) - 2 \sqrt{2}(a_3 b_3-a_2 b_4)\\
-\sqrt{6}a_2 b_1 + 2 a_4 b_4 + 3 a_1 b_2 - a_3 b_5 \\
a_2 b_2 - 3a_4 b_5 -2 a_1 b_3 + \sqrt{6} a_3 b_1\\
\end{pmatrix}
\end{math}\\  &\\

\begin{math}
\bo{3^\prime_a} \sim \begin{pmatrix}
a_1b_4 + a_2 b_3 -a_3 b_2 -a_4 b_1\\
\sqrt{2}(a_3 b_4 - a_4 b_3)\\
\sqrt{2}(a_1 b_2 -a_2 b_1)\\
\end{pmatrix}
\end{math}&

\begin{math}
\bo{4} \sim \begin{pmatrix}
\sqrt{3}a_1 b_1 + \sqrt{2}(a_3 b_4 -a_2 b_5 -2 a_4 b_3)\\
\sqrt{2}(-a_1 b_2 +a_4 b_4 + 2a_3 b_5)- \sqrt{3}a_2 b_1 \\
\sqrt{2}(a_1 b_3+ 2 a_2 b_2 -a_4 b_5 )-\sqrt{3} a_3 b_1\\
\sqrt{2}(-2a_1 b_4 + a_2 b_3 - a_3 b_2) + \sqrt{3} a_4 b_1\\
\end{pmatrix}
\end{math}\\ & \\
\begin{math}
\bo{4_s} \sim \begin{pmatrix}
a_2 b_4 + a_3 b_3 + a_4 b_2\\
a_1 b_1 + a_3 b_4 + a_4 b_3 \\
a_1 b_2+ a_2 b_1 + a_4 b_4\\
a_1 b_3 + a_3 b_1 + a_2 b_2\\
\end{pmatrix}
\end{math}&
 
\begin{math}
\bo{5_1} \sim \begin{pmatrix}
\sqrt{2}(a_1 b_5 -a_2 b_4 - a_3 b_3 + a_4 b_2)\\
-\sqrt{2}a_1 b_1 -\sqrt{3}(a_3 b_4 + a_4 b_3)\\
\sqrt{2} a_2 b_1 + \sqrt{3}(a_1 b_2 + a_3 b_5)\\
\sqrt{2} a_3 b_1 + \sqrt{3}(a_2 b_2 + a_4 b_5)\\
-\sqrt{2}a_4 b_1 -\sqrt{3}(a_1 b_4+ a_2 b_3)\\
\end{pmatrix}
\end{math}\\& \\

\begin{math}
\bo{5_s} \sim \begin{pmatrix}
\sqrt{3}(a_1 b_4 - a_2 b_3 - a_3 b_2 + a_4 b_1)\\
-\sqrt{2}(a_2 b_4+ a_4 b_2 - 2a_3 b_3)\\
\sqrt{2}(-2a_1 b_1 + a_3 b_4 + a_4 b_3)\\
\sqrt{2}(a_1 b_2 + a_2 b_1 -2 a_4 b_4)\\
\sqrt{2}(-a_1 b_3 + 2 a_2 b_2 -a_3 b_1)\\
\end{pmatrix}
\end{math} &
\begin{math}
\bo{5_2} \sim \begin{pmatrix}
2(a_1 b_5 + a_4 b_2) + 4(a_2 b_4 + a_3 b_3)\\
2(2a_1 b_1 + \sqrt{6}a_2 b_5) \\
-\sqrt{6}(a_1 b_2 +a_3 b_5 - 2 a_4 b_4) + 2 a_2 b_1\\
\sqrt{6}(2 a_1 b_3 - a_2 b_2 - a_4 b_5) + 2 a_3 b_1\\
2(\sqrt{6}a_3 b_2 + 2 a_4 b_1)\\
\end{pmatrix}
\end{math}\\& \\
\hline
\end{tabular}
\end{center}

\begin{center}
\begin{tabular}{|c|}
\hline
 $\bo{5} \ot
\bo{5} = \bo{1_s} \op \bo{3_a} \op \bo{3^\prime_a} \op \bo{4_s} \op \bo{4_a}
\op \bo{5_{1,s}} \op \bo{5_{2,s}}$ \\
\hline
\hline
\\
$\bo{1_s} \sim a_1b_1 + a_2b_5 + a_3 b_4 + a_4 b_3 + a_5 b_2$\\  \\

\begin{math}
\bo{3_a} \sim \begin{pmatrix}
a_2 b_5 - a_5 b_2 + 2 (a_3 b_4 - a_4 b_3)\\
\sqrt{3}(a_2 b_1 - a_1 b_2) + \sqrt{2}(a_3 b_5 - a_5 b_3)\\
\sqrt{3}(a_1 b_5- a_5 b_1) + \sqrt{2} (a_2 b_4 - a_4 b_2)\\
\end{pmatrix}
\end{math}\\ \\

\begin{math}
\bo{3_a^\prime} \sim \begin{pmatrix}
2 (a_2 b_5 - a_5 b_2) -  a_3 b_4 + a_4 b_3\\
\sqrt{3}(a_1 b_3 - a_3 b_1) + \sqrt{2}(a_4 b_5 - a_5 b_4)\\
- \sqrt{3}(a_1 b_4 - a_4 b_1) + \sqrt{2} (a_2 b_3 - a_3 b_2)\\
\end{pmatrix}
\end{math}\\  \\

\begin{math}
\bo{4_s} \sim \begin{pmatrix}
3 \sqrt{2}(a_1b_2 + a_2 b_1)- \sqrt{3}( a_3 b_5 - 4 a_4 b_4 + a_5 b_3)\\
3 \sqrt{2}(a_1 b_3 + a_3 b_1) - \sqrt{3}(-4 a_2 b_2 + a_4 b_5 + a_5 b_4)\\
3\sqrt{2} (a_1 b_4 +a_4 b_1) -\sqrt{3} (a_2 b_3 + a_3 b_2 - 4 a_5 b_5)\\
3 \sqrt{2} (a_1 b_5 + a_5 b_1) - \sqrt{3}(a_2 b_4 - 4 a_3 b_3 + a_4 b_2)\\
\end{pmatrix}
\end{math} \\  \\

\begin{math}
\bo{4_a} \sim \begin{pmatrix}
\sqrt{2}(a_1b_2 - a_2 b_1) + \sqrt{3}(a_3 b_5 - a_5 b_3)\\
\sqrt{2}(a_3 b_1 - a_1 b_3) + \sqrt{3}(a_4 b_5- a_5 b_4)\\
\sqrt{2}(a_4 b_1 - a_1 b_4) + \sqrt{3}(a_3 b_2 - a_2 b_3)\\
\sqrt{2}(a_1 b_5 - a_5 b_1) + \sqrt{3}(a_4 b_2 - a_2 b_4)\\
\end{pmatrix}
\end{math} \\   \\
 
\begin{math}
\bo{5_{1,s}} \sim \begin{pmatrix}
2(a_1 b_1 - a_3 b_4 - a_4 b_3) + a_2 b_5 + a_5 b_2\\
a_1 b_2 + a_2 b_1 + \sqrt{6}(a_3 b_5 + a_5 b_3)\\
\sqrt{6}a_2 b_2 - 2(a_1b_3 + a_3 b_1)\\
\sqrt{6} a_5 b_5 - 2(a_1 b_4 + a_4 b_1)\\
\sqrt{6}(a_2 b_4 + a_4 b_2) + a_1 b_5 + a_5 b_1\\
\end{pmatrix}
\end{math}\\     \\

\begin{math}
\bo{5_{2,s}} \sim \begin{pmatrix}
2(a_1 b_1 - a_2 b_5 - a_5 b_2) + a_3 b_4 + a_4 b_3\\
\sqrt{6}a_4 b_4 - 2(a_1b_2 + a_2 b_1)\\
a_1 b_3 + a_3 b_1 + \sqrt{6}(a_4 b_5 + a_5 b_4)\\
\sqrt{6}(a_2 b_3 + a_3 b_2) + a_1 b_4 + a_4 b_1\\
\sqrt{6} a_3 b_3 - 2(a_1 b_5 + a_5 b_1)\\
\end{pmatrix}
\end{math}\\  \\
\hline
\end{tabular}
\end{center}

\end{document}